\documentclass{article}

\usepackage{algorithm}
\usepackage[margin=30mm,includehead,includefoot]{geometry}
\usepackage[noend]{algpseudocode}
\usepackage{amsmath,amssymb,amsthm,amscd,color,comment}
\usepackage{autobreak}
\usepackage{color}
\usepackage{xcolor}
\usepackage{url}
\usepackage{blkarray}
\usepackage{jheppub}
\usepackage{empheq}
\usepackage{graphicx}
\usepackage{booktabs, multirow}
\usepackage{subcaption}
\usepackage{bm}
\usepackage{enumerate}
\usepackage{comment}
\usepackage{parskip}
\usepackage{bbm}
\usepackage{slashed}

\usepackage{tikz}
\usepackage[compat=1.1.0]{tikz-feynman}
\usetikzlibrary{patterns}
\usetikzlibrary{arrows,shapes,positioning}
\usetikzlibrary{trees}
\usetikzlibrary{matrix,arrows} 
\usetikzlibrary{positioning}
\usetikzlibrary{calc,through}
\usetikzlibrary{decorations.pathreplacing}
\usepackage{pgffor}
\usetikzlibrary{decorations.pathmorphing}
\usetikzlibrary{decorations.markings}
\tikzstyle{block} = [draw, rectangle, 
minimum height=3em, minimum width=6em]

\tikzstyle{decision} = [diamond, draw, fill=blue!20, 
text width=4.5em, text badly centered, node distance=3cm, inner sep=0pt]
\tikzstyle{block} = [rectangle, draw, fill=blue!10, 
text width=7em, text centered, rounded corners, minimum height=4em, node distance=3cm]
\tikzstyle{line} = [draw, -latex']
\tikzstyle{cloud} = [draw, rectangle, fill=blue!20, text width=6em, text centered, rounded corners, node distance=3cm,
minimum height=2em]
\tikzstyle{border} = [draw, dashed, rectangle, fill=blue!5, rounded corners, node distance=3cm, minimum height=25em, minimum width=22em]
\tikzstyle{data} = [draw, rectangle, fill=blue!10, rounded corners, minimum height=2em, minimum width=8em]
\tikzstyle{box} = [draw, rectangle, fill=white, rounded corners, minimum width=10em]
\tikzstyle{input}=[trapezium, draw, text centered, trapezium left angle=60, trapezium right angle=120, minimum height=2em, fill=blue!10]

\tikzset{fermionnoarrow/.style={draw=black},}

\definecolor{green1}{HTML}{3D792A}
\definecolor{cyan1}{HTML}{37cdaa}
\definecolor{blue1}{HTML}{5d7ac4}
\definecolor{red1}{HTML}{d0482a}
\definecolor{purple1}{HTML}{845ea8}
\definecolor{orange1}{HTML}{e07229}

\newcommand{\bega}{\begin{gathered}}
\newcommand{\ega}{\end{gathered}}

\title{Two-loop vertices with vacuum polarization insertion}

\author[a]{Taushif Ahmed,}
\author[b,c]{Giulio Crisanti,}
\author[d]{Federico Gasparotto,}
\author[a]{Syed Mehedi Hasan,}
\author[b,c]{Pierpaolo Mastrolia}

\affiliation[a]{\uniregensburg}
\affiliation[b]{\unipd}
\affiliation[c]{\pdinfn}
\affiliation[d]{\unimainz}

\newcommand{\uniregensburg}{Institute for Theoretical Physics, University of Regensburg, 93040 Regensburg, Germany }

\newcommand{\unipd}{Dipartimento di Fisica e Astronomia, Universit\`a degli Studi di Padova,
Via Marzolo 8, I-35131 Padova, Italy.}

\newcommand{\pdinfn}{INFN, Sezione di Padova,
Via Marzolo 8, I-35131 Padova, Italy.}

\newcommand{\unimainz}{PRISMA Cluster of Excellence, Institut f\"ur Physik, Staudinger Weg 7, Johannes Gutenberg-Universit\"at Mainz, D - 55099 Mainz, Germany}

\vspace{4cm}

\emailAdd{taushif.ahmed@ur.de}
\emailAdd{giulioeugenio.crisanti@phd.unipd.it}
\emailAdd{fgasparo@uni-mainz.de}
\emailAdd{Syed-Mehedi.Hasan@physik.uni-regensburg.de}
\emailAdd{pierpaolo.mastrolia@unipd.it}

\abstract{
We present the analytic evaluation of the second-order corrections to the massive form factors, due to two-loop vertex diagrams with a vacuum polarization insertion, with exact dependence on the 
external and internal fermion masses, and on the squared momentum transfer.
We consider vector, axial-vector, scalar and pseudoscalar interactions between the external fermion and the external field. 
After renormalization, the finite expressions of the form factors are expressed in terms of polylogarithms up to weight three.
}

\allowdisplaybreaks

\begin{document}
\addtocontents{toc}{\protect\setcounter{tocdepth}{2}}
\preprint{MITP-23-041}

\maketitle

\newcommand{\defaultscale}{0.1}
\newcommand{\vchb}[1]{
\vcenter{\hbox{#1}}
}

\newcommand{\masterintegralfamily}[1]{
\scalebox{\defaultscale}{{\scalebox{0.85}{\includegraphics[trim={0cm 0cm 0cm 0cm},clip,scale=#1]{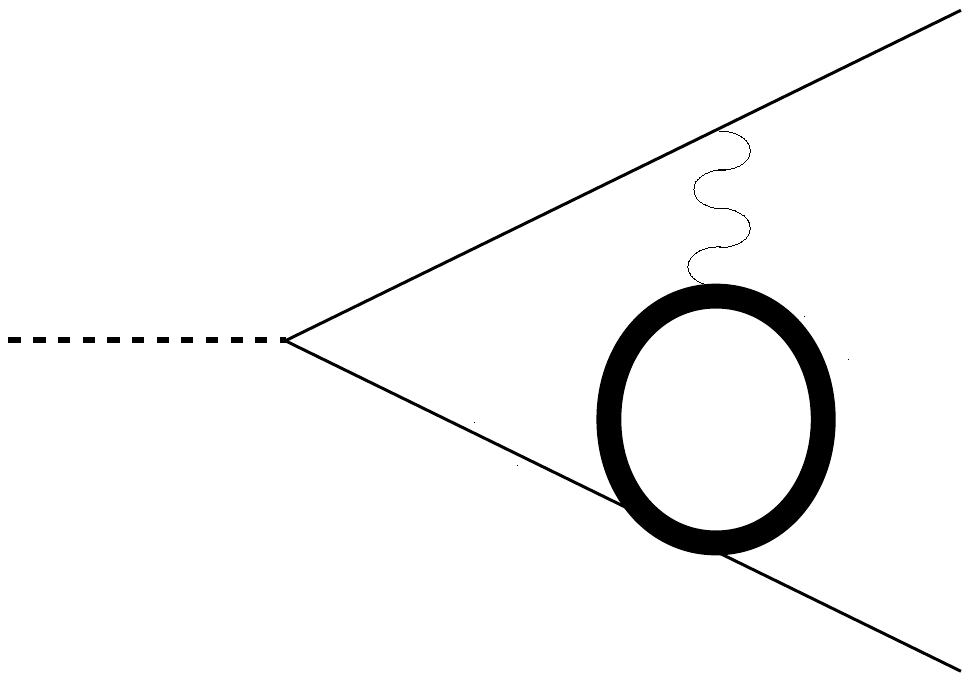}}}}
}

\newcommand{\masteri}[1]{
\scalebox{\defaultscale}{{\scalebox{0.65}{\includegraphics[trim={0cm 0cm 0cm 0cm},clip,scale=#1]{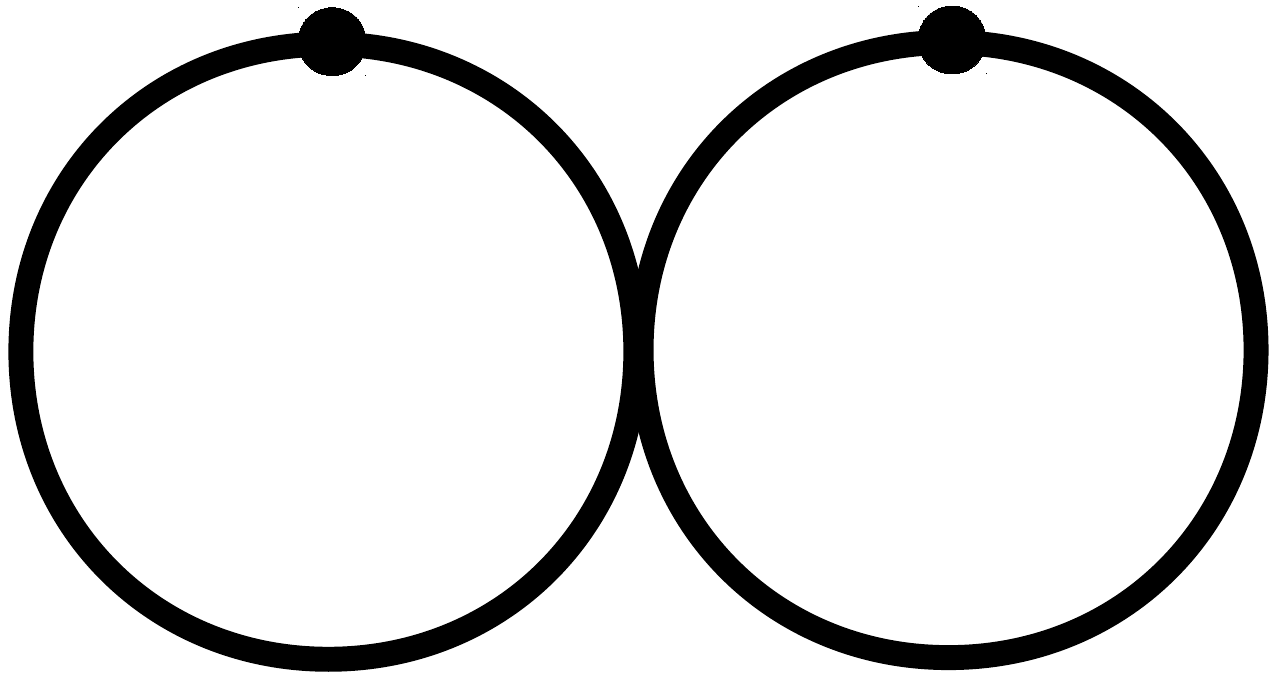}}}}
}

\newcommand{\masterii}[1]{
\scalebox{\defaultscale}{\scalebox{0.65}{\includegraphics[trim={0cm 0cm 0cm 0cm},clip,scale=#1]{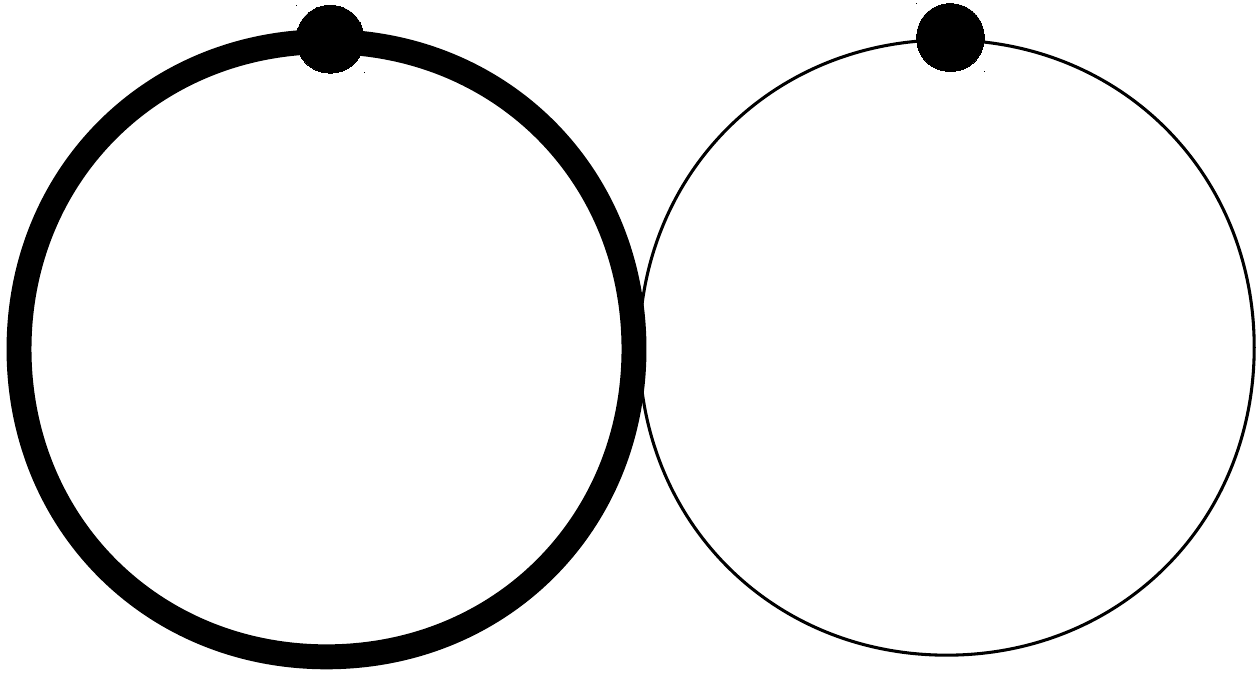}}}
}

\newcommand{\masterdotsii}[1]{
\scalebox{\defaultscale}{\scalebox{0.65}{\includegraphics[trim={3.8cm 0cm 4cm 0cm},clip,scale=#1]{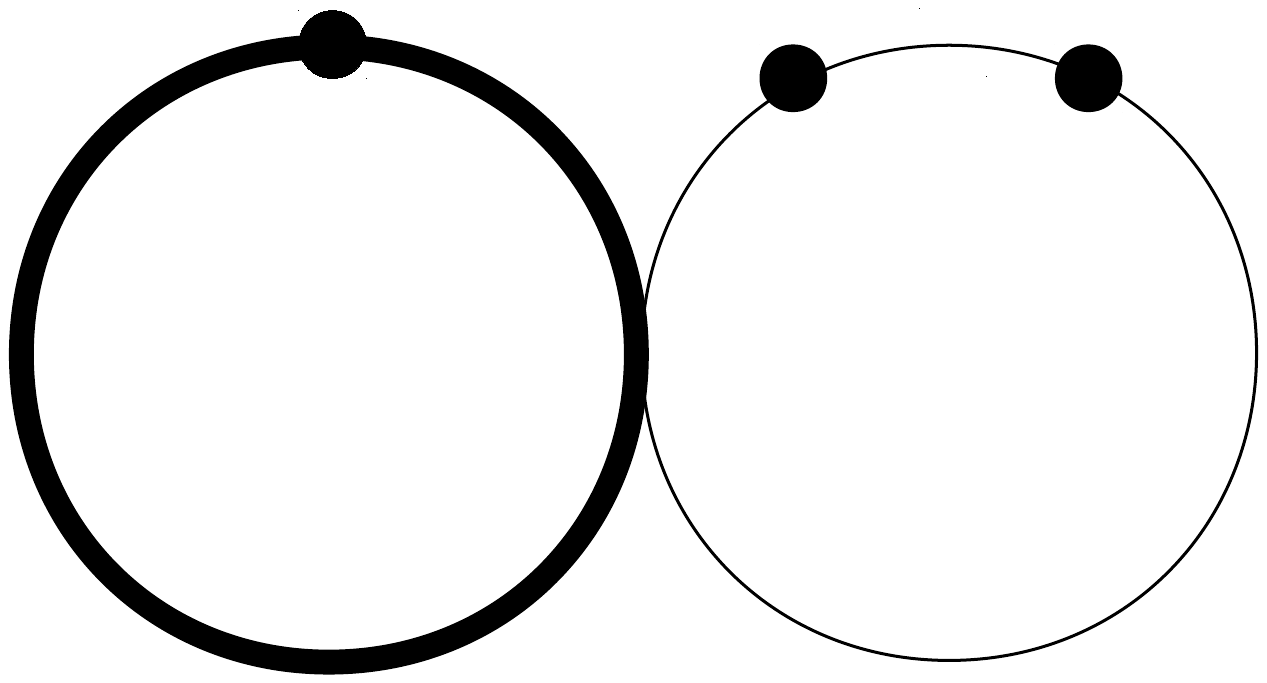}}}
}

\newcommand{\masteriii}[1]{
\scalebox{\defaultscale}{\scalebox{0.65}{\includegraphics[trim={0cm 0cm 0cm 0cm},clip,scale=#1]{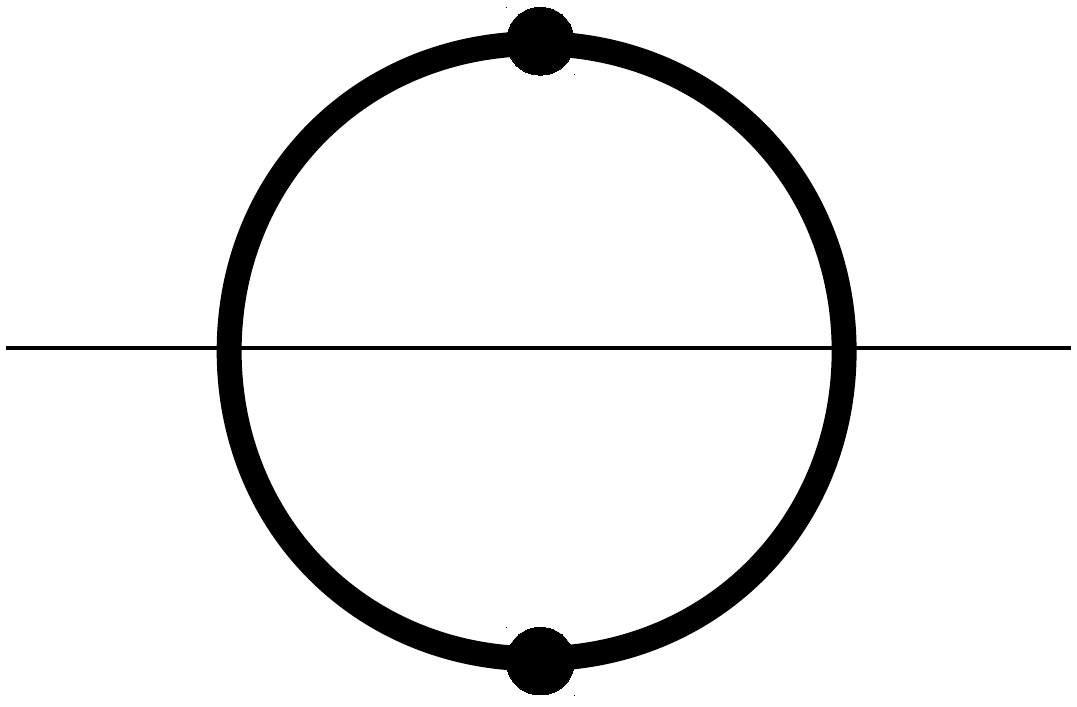}}}
}

\newcommand{\masteriv}[1]{
\scalebox{\defaultscale}{\scalebox{0.65}{\includegraphics[trim={0cm 0cm 0cm 0cm},clip,scale=#1]{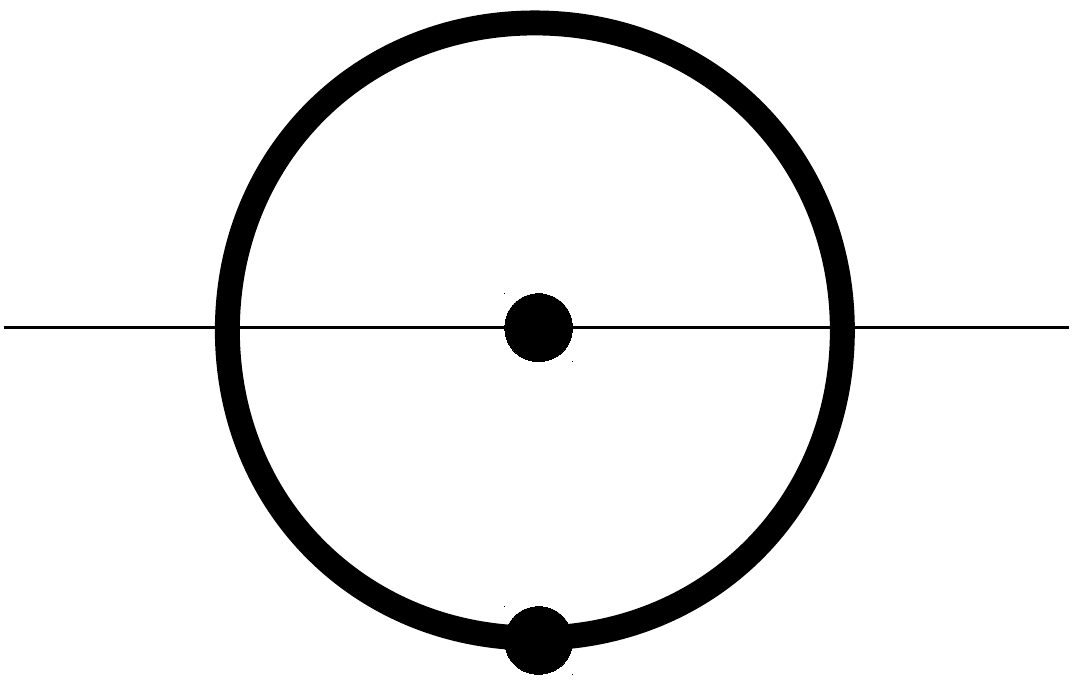}}}
}

\newcommand{\masterdotsiv}[1]{
\scalebox{\defaultscale}{\scalebox{0.65}{\includegraphics[trim={4cm 0cm 7cm 0cm},clip,scale=#1]{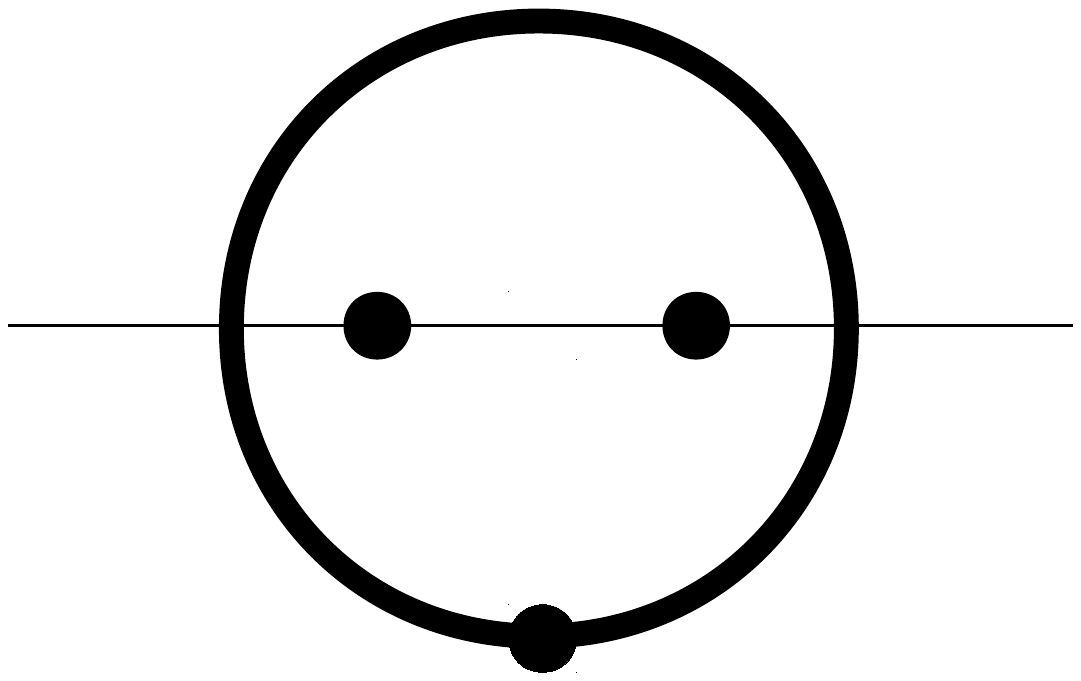}}}
}

\newcommand{\masterv}[1]{
\scalebox{\defaultscale}{\scalebox{0.6}{\includegraphics[trim={0cm 0cm 0cm 0cm},clip,scale=#1]{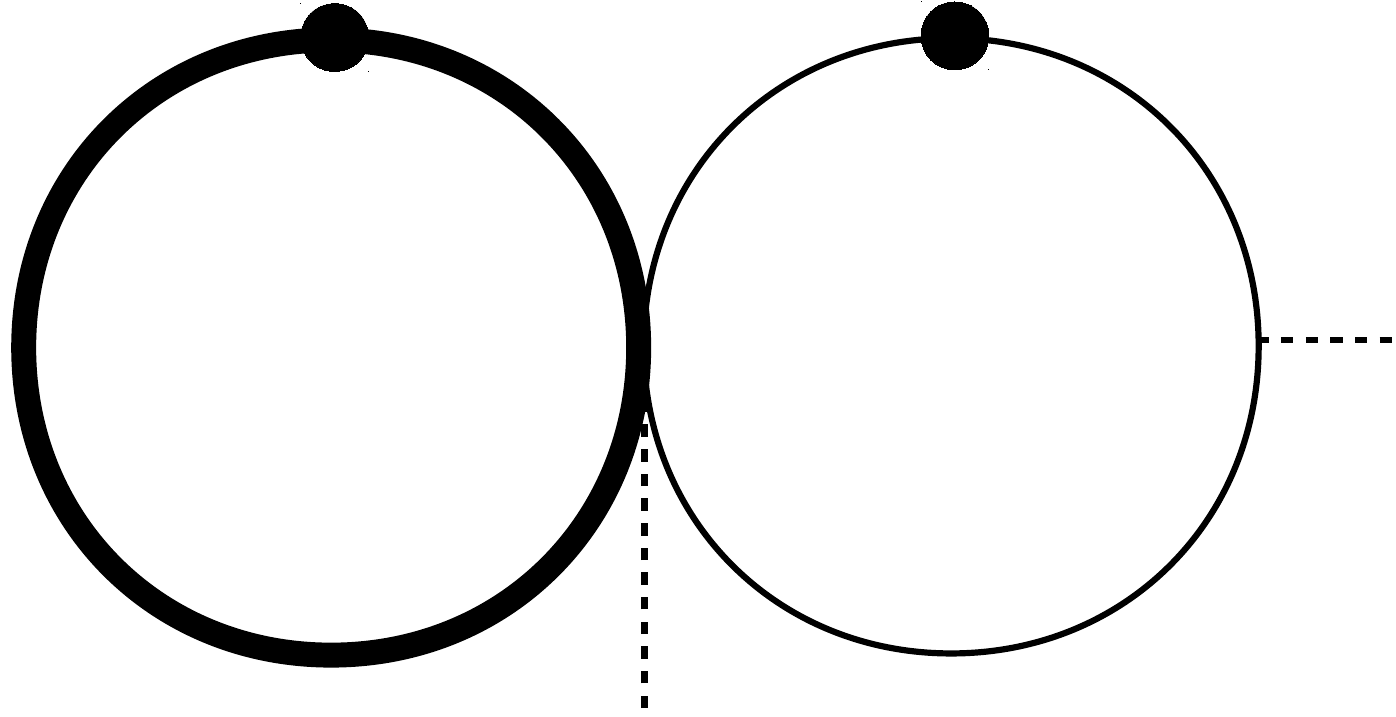}}}
}

\newcommand{\mastervi}[1]{
\scalebox{\defaultscale}{\scalebox{0.85}{\includegraphics[trim={0cm 0cm 0cm 0cm},clip,scale=#1]{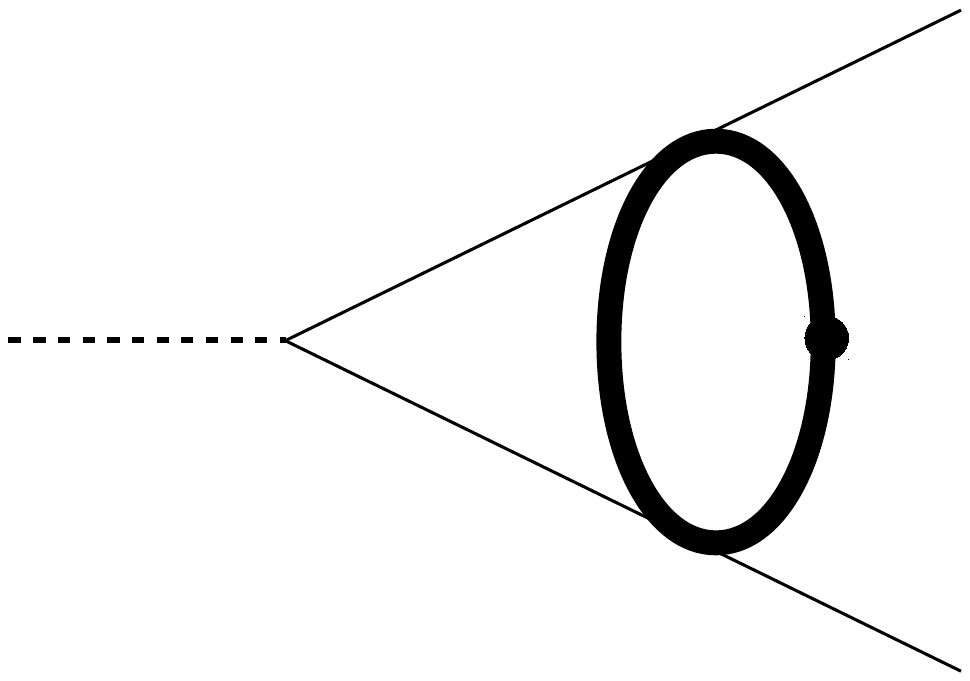}}}
}

\newcommand{\mastervii}[1]{
\scalebox{\defaultscale}{\scalebox{0.85}{\includegraphics[trim={0cm 0cm 0cm 0cm},clip,scale=#1]{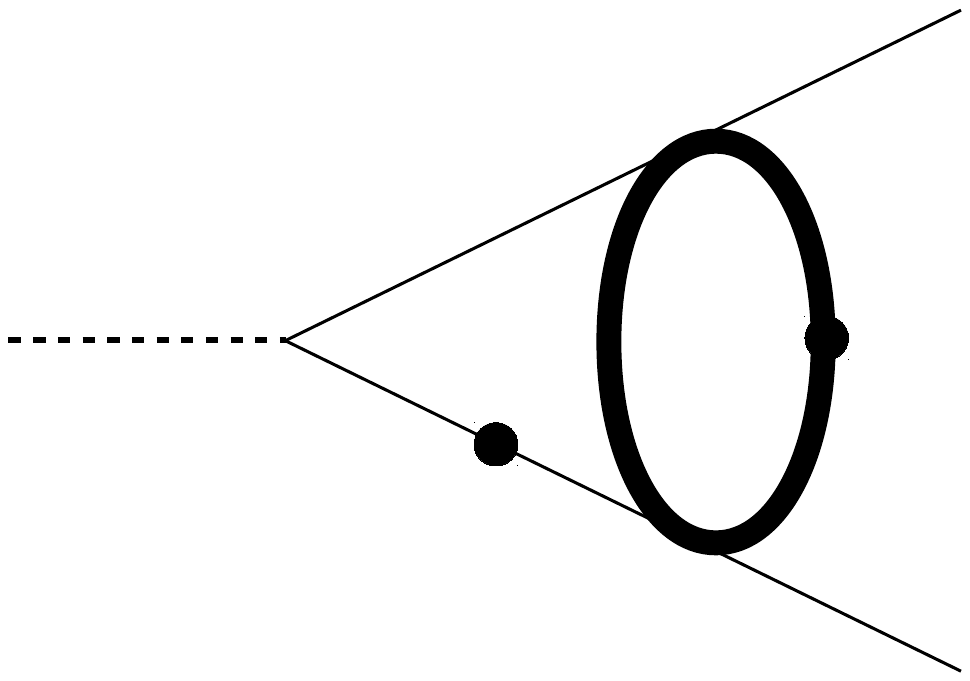}}}
}

\newcommand{\iloopmastermi}[1]{
\scalebox{\defaultscale}{\scalebox{0.6}{\includegraphics[trim={0cm 0cm 0cm 0cm},clip,scale=#1]{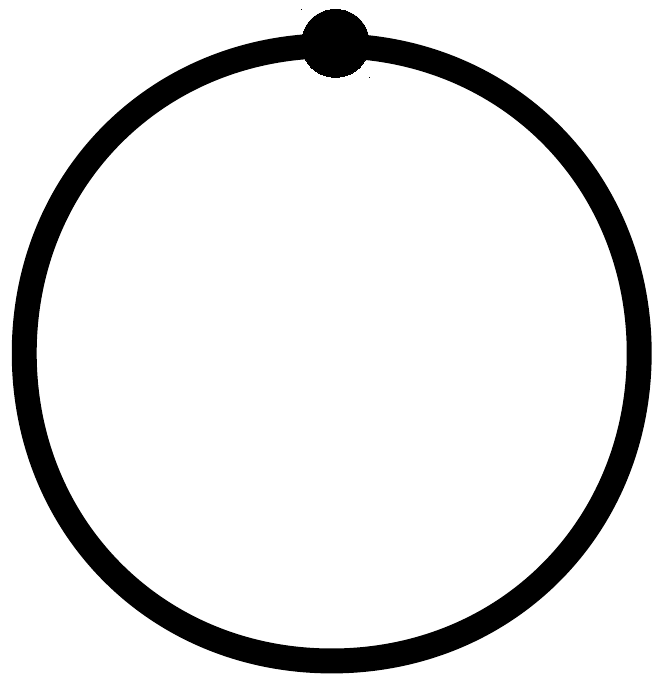}}}
}

\newcommand{\iloopmasterme}[1]{
\scalebox{\defaultscale}{\scalebox{0.6}{\includegraphics[trim={0cm 0cm 0cm 0cm},clip,scale=#1]{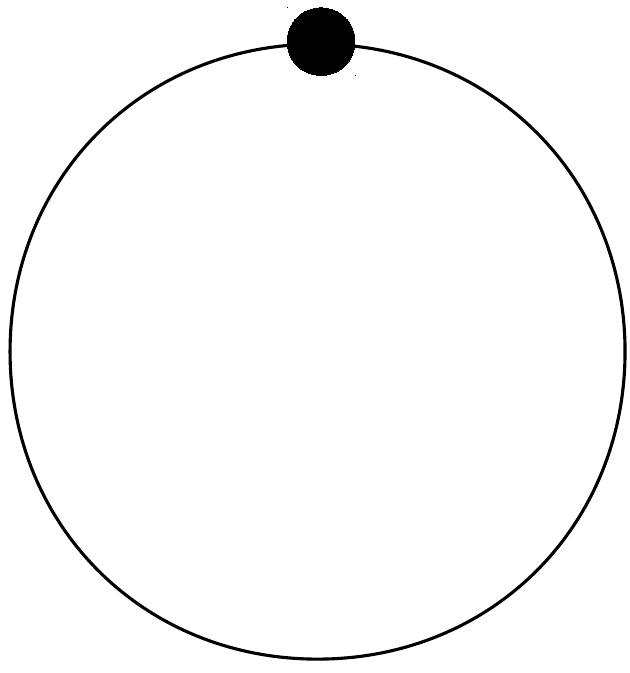}}}
}

\newcommand{\iloopscalarsunrise}[1]{
\scalebox{\defaultscale}{\scalebox{0.6}{\includegraphics[trim={0cm 0cm 0cm 0cm},clip,scale=#1]{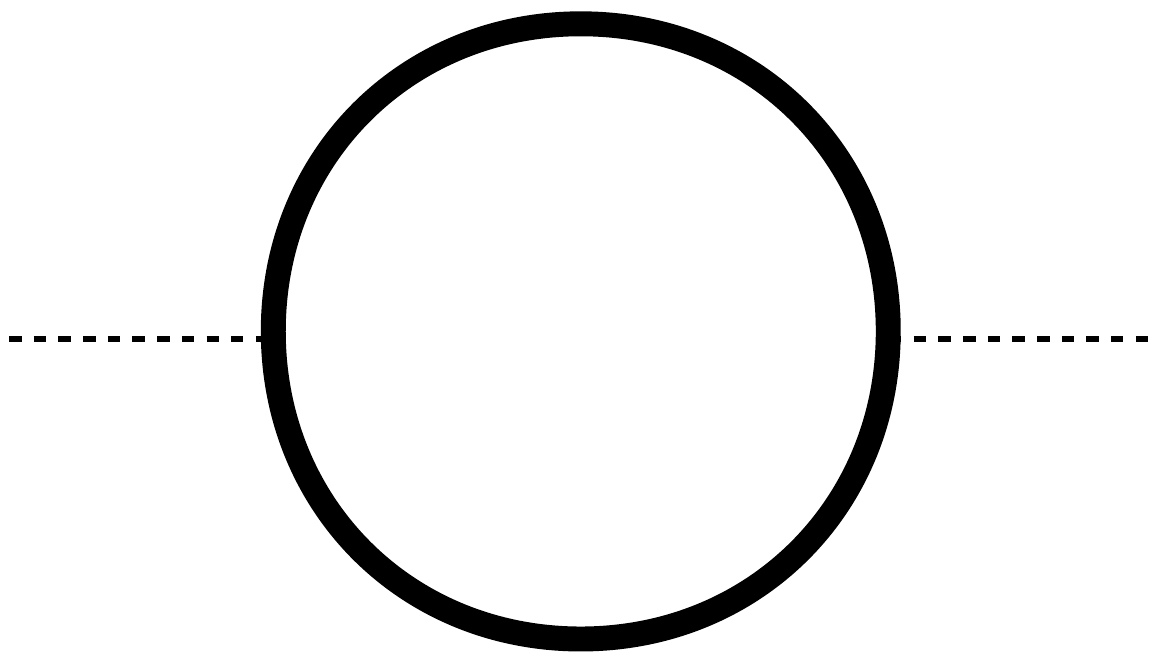}}}
}

\newcommand{\maintopology}[1]{
\scalebox{\defaultscale}{\scalebox{2}{\includegraphics[trim={0cm 0cm 0cm 0cm},clip,scale=#1]{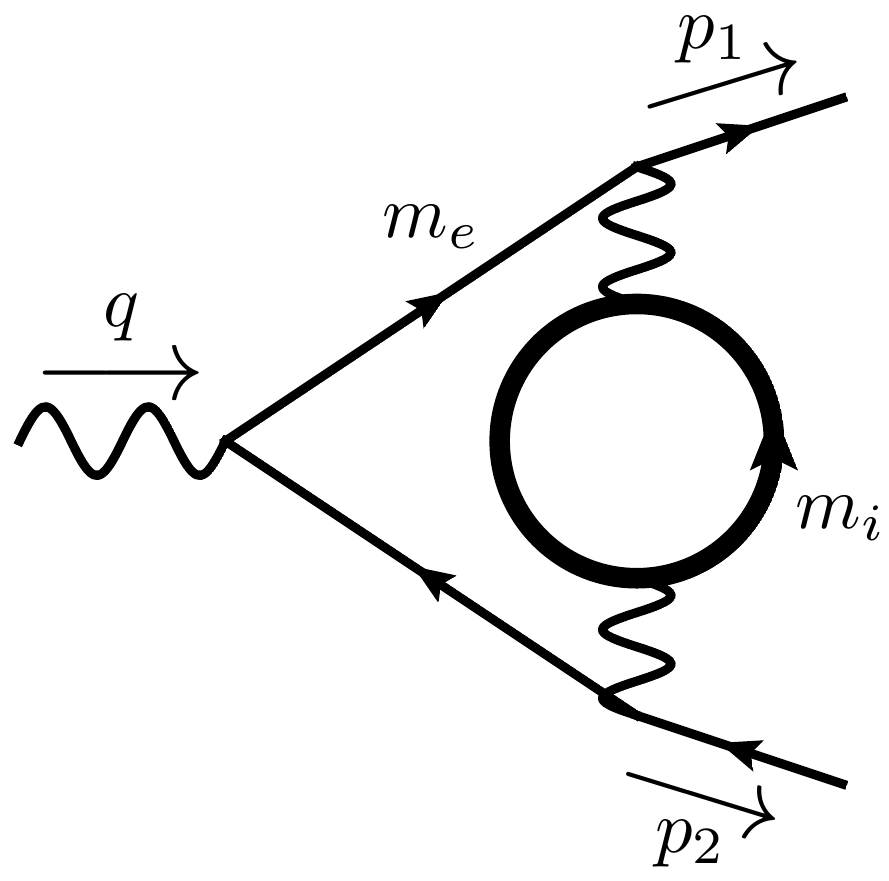}}}
}

\newcommand{\maintopologynolab}[1]{
\scalebox{\defaultscale}{\scalebox{2}{\includegraphics[trim={0cm 0cm 0cm 0cm},clip,scale=#1]{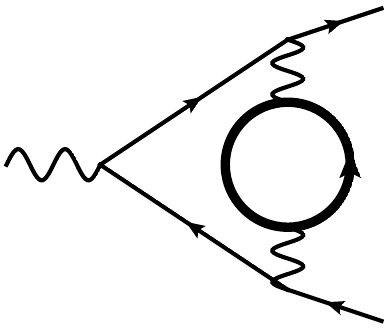}}}
}

\newcommand{\iloopspinorsunrise}[1]{
\scalebox{\defaultscale}{\scalebox{3}{\includegraphics[trim={0cm 0cm 0cm 0cm},clip,scale=#1]{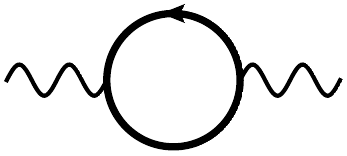}}}
}

\newcommand{\ilooptriangle}[1]{
\scalebox{\defaultscale}{\scalebox{2}{\includegraphics[trim={0cm 0cm 0cm 0cm},clip,scale=#1]{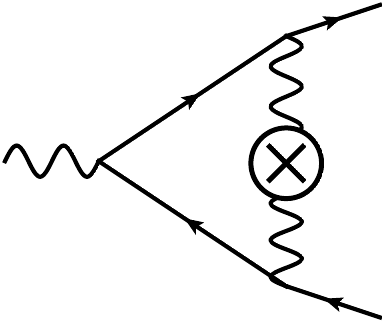}}}
}

\newcommand{\ziicountertermins}[1]{
\scalebox{\defaultscale}{\scalebox{2.5}{\includegraphics[trim={0cm 0cm 0cm 0cm},clip,scale=#1]{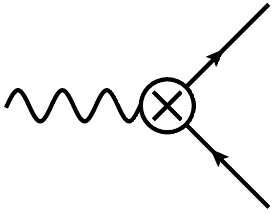}}}
}

\newcommand{\ziicounterterm}[1]{
\scalebox{\defaultscale}{\scalebox{2}{\includegraphics[trim={0cm 0cm 0cm 0cm},clip,scale=#1]{z2counterterm2.pdf}}}
}

\newcommand{\ziiicountertermins}[1]{
\scalebox{\defaultscale}{\scalebox{3}{\includegraphics[trim={0cm 0cm 0cm 0cm},clip,scale=#1]{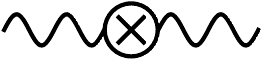}}}
}

\newcommand{\ziiicounterterm}[1]{
\scalebox{\defaultscale}{\scalebox{2}{\includegraphics[trim={0cm 0cm 0cm 0cm},clip,scale=#1]{z3counterterm2.pdf}}}
}

\newcommand{\maintopologyQCD}[1]{
\scalebox{\defaultscale}{\scalebox{2}{\includegraphics[trim={0cm 0cm 0cm 0cm},clip,scale=#1]{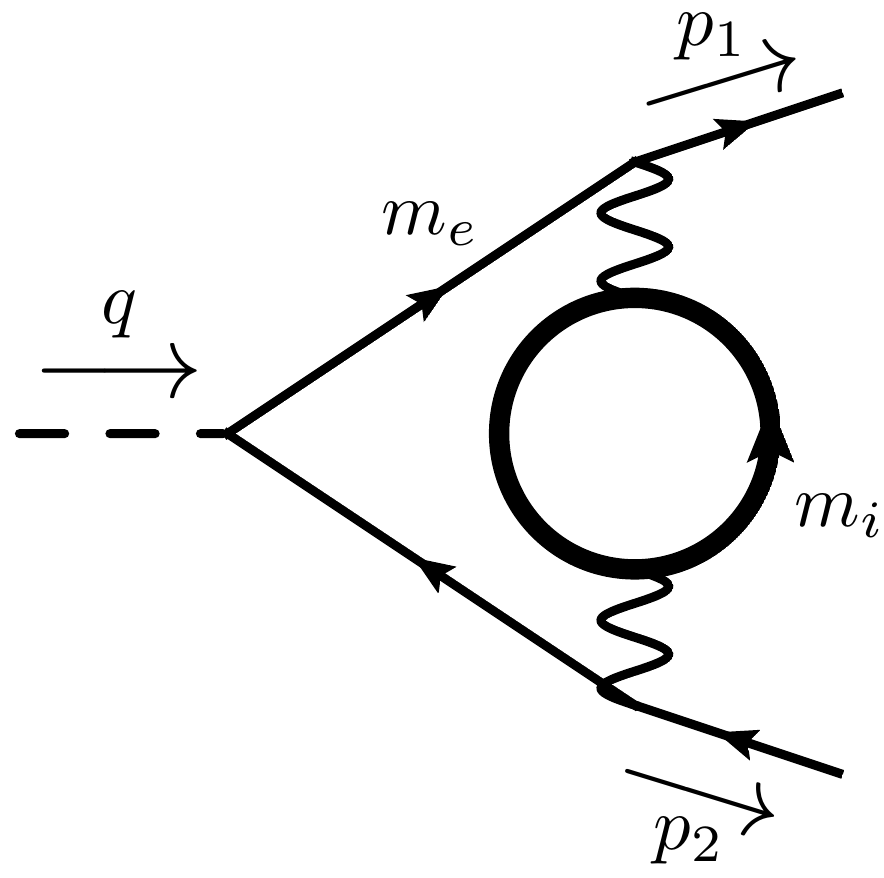}}}
}

\newcommand{\maintopologynolabQCD}[1]{
\scalebox{\defaultscale}{\scalebox{2}{\includegraphics[trim={0cm 0cm 0cm 0cm},clip,scale=#1]{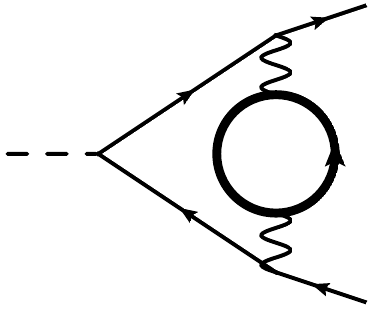}}}
}
\newcommand{\ilooptriangleQCD}[1]{
\scalebox{\defaultscale}{\scalebox{2}{\includegraphics[trim={0cm 0cm 0cm 0cm},clip,scale=#1]{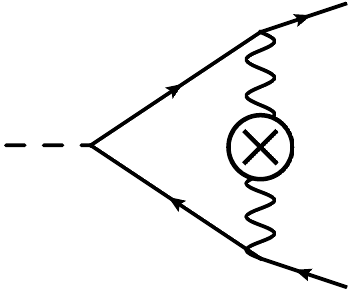}}}
}
\newcommand{\ziicounterterminsQCD}[1]{
\scalebox{\defaultscale}{\scalebox{2.5}{\includegraphics[trim={0cm 0cm 0cm 0cm},clip,scale=#1]{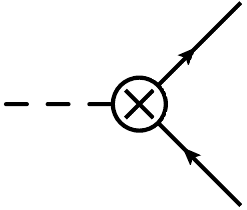}}}
}
\newcommand{\ziiicounterterminsQCD}[1]{
\scalebox{\defaultscale}{\scalebox{3}{\includegraphics[trim={0cm 0cm 0cm 0cm},clip,scale=#1]{z3countertermQCD.pdf}}}
}
\newcommand{\iloopspinorsunriseQCD}[1]{
\scalebox{\defaultscale}{\scalebox{3}{\includegraphics[trim={0cm 0cm 0cm 0cm},clip,scale=#1]{one loop spinor sunrise QCD.pdf}}}
}
\section{Introduction}

The analytic evaluation of the two-loop corrections to the electron form factors in Quantum Electrodynamics (QED) can be considered among the pioneering projects that triggered the developments of mathematical techniques and concepts for the evaluation of multi-loop Feynman integrals in Quantum Field Theory, which is still ongoing nowadays. 
Originally addressed by means of dispersion relations and giving a finite mass to the photon, in order to regulate the otherwise divergent integrals  \cite{Barbieri:1972as,Barbieri:1972hn,Mastrolia:2003yz},
the contributing vertex graphs were also later evaluated \cite{Bonciani:2003te,Bonciani:2003ai} within the dimensional regularization scheme, by using the differential equations method \cite{Kotikov:1990kg,Remiddi:1997ny,Gehrmann:1999as}.
The results of \cite{Mastrolia:2003yz,Bonciani:2003te,Bonciani:2003ai} were among the first studies showing that classical polylogarithms could not represent an exhaustive set of functions for Feynman integrals beyond one-loop, therefore, pointing to the need of introducing an extended set of polylogarithms, such as the Harmonic Polylogarithms (HPLs) \cite{Remiddi:1999ew} - later embedded in the wider class of Generalised Polylogarithms (GPLs) \cite{Goncharov:2001iea}. 
HPLs turned out to be useful both for the direct integration of the dispersive integrals \cite{Mastrolia:2003yz}, and for solving the differential equations of the master integrals (MIs) in terms of iterated integrals \cite{Bonciani:2003te}, in combination with suitable change of variables, needed to rationalize the integration kernels$-$a procedure that would have been later dubbed as alphabet rationalization.

The feasibility of the analytic evaluation of the electron form factors at two-loop and beyond in QED was the basis of successive studies involving the heavy-quark form factors at two-loop in Quantum Chromodynamics (QCD), in the case of a vector interaction, and
later extended, to account for the scalar, pseudoscalar, and pseudo-vector interactions 
\cite{Kniehl:1989kz,Kniehl:1994vq,Bernreuther:2004ih,Bernreuther:2004th,Bernreuther:2005rw,Bernreuther:2005gq,Bernreuther:2005gw,Czakon:2004wm,Gluza:2009yy,Ahmed:2017gyt,Ablinger:2017hst,Primo:2018zby,Lee:2018rgs}. 
From a more formal point of view, form factors turn out to be also important for investigating the singular behavior of massive amplitudes in gauge theories, through their relation to the corresponding massless approximation \cite{Mitov:2006xs}. 

In the context of scattering processes, the same two-loop vertex diagrams considered in the above studies appear, on the one side, among the (factorised) diagrams contributing to processes with the same number of external particles, yet with a larger number of loops \cite{Ablinger:2018yae,Blumlein:2019oas,Fael:2022rgm,Fael:2022miw,Fael:2023zqr,Blumlein:2023uuq},
and, on the other side, to processes having the same number of loops, yet with more legs. In this respect, the two-loop three-point integrals contributing to the vector form factors also appear in the evaluation of the two-loop QED corrections to the amplitude of the four-fermion scattering~\cite{Budassi:2021twh,Bonciani:2021okt,Broggio:2022htr} and in the two-loop QCD corrections to heavy-quark pair production in the light-quark fusion channel~\cite{Mandal:2022vju}.

In this work, we present, for the first time, 
the contribution to the vertex form factors 
of a heavy lepton with mass $m_e$, coupled to a generic external particle, 
{\it i.e.} through a vector, axial-vector, scalar and axial couplings,
coming from a two-loop graph with the insertion of a (vector) gauge boson vacuum polarization, with a closed-loop of a different type of heavy lepton, with mass $m_i \neq m_e$. 

We hereby address the evaluation of the renormalised form factors 
by decomposing them, via integration-by-parts identities (IBPs) \cite{Tkachov:1981wb,Chetyrkin:1981qh} and Laporta's algorithm \cite{Laporta:2000dsw}, to a linear combination of seven master integrals, and by evaluating the latter using the Magnus method for differential equations~\cite{Argeri:2014qva}. 

After UV-renormalization, the renormalized form factors are finite and carry the complete dependence on the squared transferred-momentum $q^2 = s$, as well as on both the internal and the external lepton masses, respectively $m_i$ and $m_e$. They are expressed in terms of GPLs up to weight three; equivalent expressions in terms of classical polylogarithms are given as well. 

In the case of vector coupling,
the evaluation of the diagram considered in the current work was previously discussed in \cite{Fael:2018dmz,Fael:2019nsf}, where semi-analytic expressions of the form factors were
given as one-fold integrals of kernels that are computed analytically by means of the hyper-spherical integration method.
The numerical evaluation of our analytic results, by means of \cite{Vollinga_2004sn}, 
is in perfect agreement with the results of \cite{Fael:2018dmz} implemented in \cite{Banerjee:2020rww}. 
An independent evaluation of the same diagram was also considered in \cite{Budassi:2021twh}, where the MIs were also computed using the method of differential equations. Our calculation is performed using a different change of variables, yielding a simple structure of the system of differential equations. 
The set of MIs presented in this work have been successfully checked against the numerical values provided by {\sc SecDec} \cite{Borowka_2018} and {\sc AMFLow} \cite{Liu_2023} and are in full agreement with those presented in \cite{Budassi:2021twh}.
We also verify that the vector form factor $F_2$ at zero momentum transfer 
agrees with the known expression of the anomalous magnetic moment given in \cite{Passera:2004bj,Henn_2021}.

The analytic expression of the vector form factors presented in this work can be directly applied in updating the analyses of the next-to-next-to-leading order (NNLO) QED corrections to the four fermion scattering with two massive lepton species.
Additionally, 
they can be used also to complete the analytic evaluation of the QCD corrections to the heavy-quark form factors \cite{Bernreuther:2004ih,Bernreuther:2004th,Bernreuther:2005rw,Bernreuther:2005gq,Bernreuther:2005gw}, 
as well as the QCD corrections to the Higgs boson decay in a pair of $b{\bar b}$-quarks \cite{Primo:2018zby}.

The paper is organized as follows.
In Section \ref{QED and QCD diagrams}, the definition of the vertex function is introduced, and the corresponding form factors for the vector, axial vector, scalar and pseudoscalar are discussed in Section \ref{sec:form factors}.
In Section \ref{sec_setup}, we discuss the integral decomposition and the evaluation of the master integrals.  
In Section \ref{sec:renormalisation}, we discuss the renormalization procedure. Finally, the results and conclusions are presented in Sections \ref{sec:results} and \ref{sec:conclusion} respectively.
Appendix \ref{appendix:matrices_canonical} contains further details on the structure of the matrices appearing in the system of differential equations obeyed by the master integrals.
In Appendix \ref{appendix:renorm}, we elaborate on the renormalization and on the evaluation of the one- and two-loop counterterm diagrams and of the relative renormalization constants.

\section{Vertex diagrams with vacuum polarisation insertion}\label{QED and QCD diagrams}

\begin{figure}[htb]
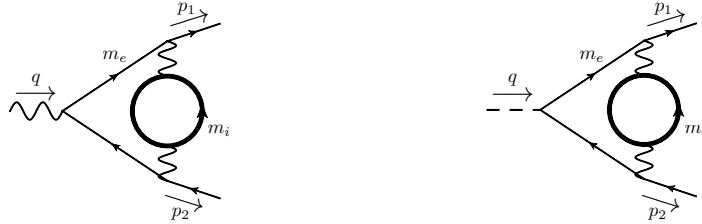

    \centering
    \begin{subfigure}{0.4\textwidth}
    \centering
    \maintopology{1}
    \end{subfigure}
    \begin{subfigure}{0.4\textwidth}
    \centering
    \maintopologyQCD{1}
    \end{subfigure}
    \caption{Two-loop vertex diagrams with vacuum polarization insertion. The left panel corresponds to the (axial-)vector case. The right panel corresponds to the (pseudo-)scalar case. The internal fermion with mass $m_i$ is represented with a thick line, while the external fermion with mass $m_e$ is represented with a thin line.
    }
    \label{fig:VkVertex}
\end{figure}

We consider the two-loop vertex diagrams $V^{(k)}$ (the index $(k)$ refers to a labelling introduced earlier in the literature \cite{Bonciani:2003te,Bonciani:2003ai,Bernreuther:2004ih,Bernreuther:2004th,Bernreuther:2005rw,Bernreuther:2005gq,Bernreuther:2005gw})
pertaining to a generic (axial-)vector or (pseudo-)scalar boson of momentum $q^\mu$ with virtuality $q^2=s$ that couples to an external on-shell fermion-antifermion pair, respectively carrying momenta $p_1$ and $p_2$ with $p_1^2=p_2^2=m_e^2$. The diagrams are subject to correction through the insertion of a vacuum polarization involving a massless vector boson coupled to a fermion-antifermion pair of mass $m_i \ne m_e$, as depicted in Fig.~\ref{fig:VkVertex}.

The  corresponding expressions are given by 
\begin{align}
    \label{fullspinstructureALL}
        V^{(k)}(s,p_1,p_2)
        = \overline{u}(p_1) \, 
        \Gamma^{(k)} \, 
        v(p_2)
\end{align}
with
\begin{align}
\label{eq:GammaALL_vector}  
        \Gamma^{(k)} &=
     \left( \frac{\alpha}{\pi} \right)^2 \times \, 
        (\mu^2)^{2\epsilon} 
        \int \left(\prod_{i=1}^2 {d^d  k_i \over (2 \pi)^{d-2}}\right)
        (-i \gamma _{\nu }) \nonumber \\ 
        &\times iS(k_1,m_e) 
        \times {\bf J}
        \times \,iS(k_1-p_1-p_2,m_e) \nonumber \\ 
        &\times (-i \gamma_{\rho}) \,
         iP^{\nu\eta}(k_1-p_1)iP^{\rho\lambda}(k_1-p_1) \nonumber \\
        &\times (-i)^2 \, (-1) \, \text{Tr}(\gamma_\eta iS(k_2,m_i)\gamma_\lambda iS(k_1+k_2-p_1,m_i)), 
    \end{align}
where: $\alpha$ is the fine structure constant, $\mu^2$ is the mass-scale introduced to keep the fine structure constant dimensionless in dimensional regularization, with $d=4-2 \epsilon$, and $k_i (i = 1,2)$ are the loop momenta. The fermion and the gauge boson propagators (in Feynman gauge) are respectively defined as
\begin{equation}
        S(q,m)=\frac{\slashed{q}+m}{q^2-m^2 + i \varepsilon}
        \quad {\rm and } \quad  
        P_{\mu\nu}(q)=-\frac{g_{\mu\nu}}{q^2 + i \varepsilon} 
\end{equation}
with the Minkowski metric $g_{\mu\nu}$.
The symbol $\mathbf{J}$ depends on the type of external boson, which is given by 
 
\begin{equation}
    \mathbf{J}= 
    \begin{cases}
         - i \left( g_{V}\gamma_{\mu} + g_{A} \gamma_5 \gamma_{\mu}  \right) \ , \quad &\text{(axial-)vector coupling}, \\
         - i \left( g_S + g_P \gamma_5 \right) \ , \quad &\text{(pseudo-)scalar coupling}.
    \end{cases}
\end{equation}
The quantities $g_V,\,g_A,\,g_S,\,g_P$ represent the vector, axial vector, scalar and pseudo scalar coupling constants respectively.

\section{Form Factors}\label{sec:form factors}
Following \cite{Bonciani:2003ai, Bernreuther:2004th,Bernreuther:2004ih}, a vertex $\Gamma_\mu$ describing the (axial-)vector coupling of fermions to a gauge boson admits a decomposition in terms of \emph{four} scalar form factors $F_{1,2}$ and $G_{1,2}$ (valid at \emph{any} number of loop and \emph{any} topology).
Depending on whether we consider (axial-)vector or (pseudo-)scalar coupling, the number of form factors changes.
For the former case, although the general decomposition involves six form factors, two do not survive due to gauge invariance. Hence, the decomposition in terms of four form factors $F_{1,2}$ and $G_{1,2}$ reads~\cite{Bonciani:2003ai, Bernreuther:2004th,Bernreuther:2004ih},
\begin{equation}
\label{ffdefs}
    \begin{split}
        \Gamma_\mu = 
        -i & \left( g_V F_1(s,m_i,m_e)\gamma_\mu
        +i{ g_V \over {2m_e}} F_2(s,m_i,m_e)
        \sigma_{\mu \alpha} q^{\alpha} \right.  \\ 
         & \left.+ g_A G_1(s,m_i,m_e) \gamma_5 \gamma_{\mu} + \frac{g_A}{2 m_e} G_2(s,m_i,m_e) \gamma_5 q_{\mu} \right) \ ,
    \end{split}
\end{equation}
with $\sigma_{\alpha\beta}=\frac{i}{2}[\gamma_\alpha,\gamma_\beta]$ and $q^{\beta}=(p_1+p_2)^{\beta}$.\\ 
~\\
The form factors $F_i$ and $G_i$ (with $i=1,2$) can be extracted from $\Gamma_\mu$ by means of suitable projectors, $P_{Fi}^\mu$ and $P_{Gi}^\mu$ as 
\begin{equation}
\label{eq:F1F2}
    F_i =\text{Tr}(P_{Fi}^\mu \Gamma_\mu )\ ,
     \qquad   {\rm and} \qquad
    G_i = \text{Tr}(P_{Gi}^\mu \Gamma_\mu )\ .
\end{equation}
The explicit expressions for the projectors in $d=4-2\epsilon$ dimensions are given by
    \begin{equation}
    \begin{split}
        P_{Fi}^\mu(s,p_1,p_2) & =
        \frac{\slashed{p}_2-m_e}{m_e}
        \,i\left(
        \,f_{i,1} \, \gamma^\mu
        +
        \frac{1}{2m_e}f_{i,2}\, (p_2-p_1)^\mu
        \right)
        \frac{\slashed{p}_1+m_e}{m_e}\ ,\\
        P_{Gi}^\mu(s,p_1,p_2) & =
        \frac{\slashed{p}_2-m_e}{m_e}
        i\,\gamma_5\left(
        g_{i,1}\gamma^\mu-\frac{1}{m_e}g_{i,2}\,(p_2+p_1)^\mu
        \right)
        \frac{\slashed{p}_1+m_e}{m_e} \ ,
        \end{split}
    \end{equation}
with
\begin{eqnarray}
f_{1,1} &=& \frac{m_e^2}{2g_V(d-2)(4m_e^2-s)} \ , \qquad 
f_{1,2}  = \frac{2(d-1)m_e^4}{g_V(d-2)(4m_e^2-s)^2} \ , \\
f_{2,1} &=& -\frac{2 m_e^4}{g_V(d-2)s(4m_e^2-s)} \ , \qquad 
f_{2,2}  = -\frac{2 m_e^4(4m_e^2+(d-2)s)}{g_V(d-2)s(4m_e^2-s)^2} \ ,
\end{eqnarray}
and 
\begin{eqnarray}
g_{1,1} &=&  \frac{m_e^2}{2g_A(d-2)(4m_e^2-s)} \ , \qquad 
g_{1,2}  =  \frac{m_e^4}{g_A(d-2)s(4m_e^2-s)},  \\
g_{2,1} &=&  \frac{2m_e^4}{g_A(d-2)s(4m_e^2-s)} \ , \quad
g_{2,2}  =  \frac{4 (d-1)\,m_e^6-(d-2)\,sm_e^4}{g_A(d-2) s^2 (4m_e^2-s)} \ .
\end{eqnarray}

In the derivation of the aforementioned projectors $P_{Gi}^\mu(s,p_1,p_2)$, we assume an anticommuting $\gamma_5$ in $d$ dimensions. However, we employ a non-anticommuting $\gamma_5$, as introduced by 't Hooft-Veltman~\cite{tHooft:1972tcz} and Breitenlohner-Maison~\cite{Breitenlohner:1977hr} for the computation of form factors in eq.~\eqref{eq:F1F2}. The latter is defined as
\begin{align}
    \gamma_5 = -\frac{i}{4!} \epsilon^{\mu\nu\rho\sigma} \gamma_\mu \gamma_\nu \gamma_\rho \gamma_\sigma\,.
\end{align}
We treat the Levi-Civita symbol following the prescription in~\cite{Larin:1991tj,Larin:1993tq}. In particular, the contraction of the $\epsilon^{\mu\nu\rho\sigma}$ with the one from the projector is done according to the usual mathematical identity in four dimensions, but with the Lorentz
indices of the resulting spacetime metric tensors all taken as $d$-dimensional. This is often known as Larin's prescription. 
As the Feynman diagrams considered in this article do not exhibit any anomalous behaviour$-$or, in other words, they fulfil non-anomalous Ward identities$-$the finite remainder is guaranteed to be independent of the prescription (commuting or anticommuting) adopted for $\gamma_5$~\cite{Chen:2019wyb,Ahmed:2019udm}\footnote{The form factors were also computed assuming a naive anticommuting $\gamma_5$$-$as implemented, for example, in \textsc{Package-X}$-$ finding the same result regardless of the prescription employed.}.

The vertex function for the (pseudo-)scalar admits a decomposition in terms of two scalar form factors $F_{S}$ and $F_{P}$~\cite{Bernreuther:2005gw} as
\begin{equation}
\label{eq:(pseudo)scalar_vertex}
        \Gamma = 
        -i \bigg( g_S F_S(s, m_i, m_e) + g_P \gamma_5 F_P(s, m_i, m_e) \bigg)
\end{equation}
with $g_S,\,g_P$ the scalar and pseudoscalar couplings, respectively. The form factors $F_S$ and $F_P$ can be  extracted from $\Gamma$ by means of suitable projectors, $P_{S}$ and $P_{P}$, as 
\begin{equation}
\label{eq:FSFP}
    F_S =\text{Tr}(P_{S} \Gamma ) \ ,
     \qquad   
    F_P = \text{Tr}(P_{P} \Gamma ) \ ,
\end{equation}
where the expressions for the projectors are given by
    \begin{equation}
    \begin{split}
        P_{S}(s,p_1,p_2) & =  \frac{\slashed{p}_2-m_e}{m_e}
        \left(
        \frac{i\,m_e^2}{2g_S(s-4m_{e}^{2})}
        \right)
        \frac{\slashed{p}_1+m_e}{m_e} \ ,
       \\
        P_{P}(s,p_1,p_2) & = \frac{\slashed{p}_2-m_e}{m_e}
        \left(
        \frac{-i\,m_e^2}{2g_{P}s}\gamma_5
        \right)
        \frac{\slashed{p}_1+m_e}{m_e} \ ,
        \end{split}
    \end{equation}

The form factors $F_i$ and $G_i$ can be computed perturbatively, as series expansions in powers of $(\alpha/\pi)$, as
\begin{equation}
\begin{split}
F_{1}&=1+\left(\frac{\alpha}{\pi}\right)F_{1}^{(1)}+\left(\frac{\alpha}{\pi}\right)^2F_{1}^{(2)}+\mathcal{O}\left(\frac{\alpha}{\pi}\right)^3, \qquad F_{2}=\left(\frac{\alpha}{\pi}\right)F_{2}^{(1)}+\left(\frac{\alpha}{\pi}\right)^2F_{2}^{(2)}+\mathcal{O}\left(\frac{\alpha}{\pi}\right)^3,\\
G_{1}&=1+\left(\frac{\alpha}{\pi}\right)G_{1}^{(1)}+\left(\frac{\alpha}{\pi}\right)^2G_{1}^{(2)}+\mathcal{O}\left(\frac{\alpha}{\pi}\right)^3, 
\qquad
G_{2}=\left(\frac{\alpha}{\pi}\right)G_{2}^{(1)}+\left(\frac{\alpha}{\pi}\right)^2G_{2}^{(2)}+\mathcal{O}\left(\frac{\alpha}{\pi}\right)^3,\\
F_{S}&=1+\left(\frac{\alpha}{\pi}\right)F_{S}^{(1)}+\left(\frac{\alpha}{\pi}\right)^2F_{S}^{(2)}+\mathcal{O}\left(\frac{\alpha}{\pi}\right)^3,\\
F_{P}&=1+\left(\frac{\alpha}{\pi}\right)F_{P}^{(1)}+\left(\frac{\alpha}{\pi}\right)^2F_{P}^{(2)}+\mathcal{O}\left(\frac{\alpha}{\pi}\right)^3 \ ,
\end{split}
\end{equation}
where the superscripts $(1)$ and $(2)$ indicate the number of loops of the contributing diagrams. 
The analytic evaluation of the contributions of the two-loop diagrams in Figure \ref{fig:VkVertex} to the form factors $F_i^{(2)}$ and $G_i^{(2)}$, 
keeping complete dependence on the masses of the internal and of the external fermions is the main result of this work. We denote these contributions as $F_{i}^{(k)}$ and $G_{i}^{(k)}$.

\section{Computation of Form Factors}\label{sec_setup}
The evaluation of the form factors $F_{i}^{(k)}$ and $G_{i}^{(k)}$ proceeds by applying the relevant projectors, defined in eq.~(\ref{eq:GammaALL_vector}) to the vertices $\Gamma_{\mu}^{(k)}$ and $\Gamma^{(k)}$. To evaluate the projectors, the Lorentz and Dirac algebra is performed in $d$ dimensions and implemented in the Mathematica packages \textsc{Package-X}~\cite{Patel:2015tea} and \textsc{FeynCalc} \cite{MERTIG1991345} independently. The result of this operation is a linear combination of scalar Feynman integrals, all members of the same integral family given by
\begin{equation}
I_{n_1,\ldots,n_7} \equiv 
    \int \, \widetilde{d^{d}k_{1}} \, \widetilde{d^{d}k_{2}}\frac{1}{D_{1}^{n_{1}} \cdots D_{7}^{n_{7}}} \ ,
\label{eq:integralfamily}
\end{equation}

\begin{figure}[t]
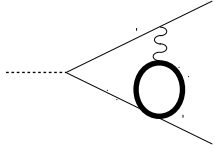

    \centering
    \masterintegralfamily{2}
\caption{Integral family associated to eq.~(\ref{eq:integralfamily}). The dashed lines denotes the external leg with momentum $q^{\mu}$, and $q^2=s$. Straight thin lines (resp. straight thick lines) denote denominators with mass $m_e$ (resp. $m_i$). Wavy lines denote massless denominators.
}
\label{fig:intfamily}
\end{figure}

where
\begin{eqnarray}
    && D_1=k_1^2-m_e^2  \ , \quad
       D_2= (k_1-p_1)^2 \ , \quad
       D_3=k_2^2-m_i^2  \ , \nonumber \\
    && D_4=(k_1+k_2-p_1)^2-m_i^2 \ , \quad
       D_5=(k_1-p_1-p_2)^2-m_e^2 \ , \nonumber \\
    && D_6= k_1 \cdot k_2 \ , \quad
       D_7= p_2 \cdot k_2 \ ,
\end{eqnarray}
with
$p_1\cdot p_2=(s-2m_e^2)/2,\,p_1^2=p_2^2=m_e^2$,
represented in Figure~\ref{fig:intfamily}.
We observe that $D_1,\ldots,D_5$ carry the momentum flowing through the diagram propagators, while $D_6$ and $D_7$ are auxiliary denominators related to irreducible scalar products.

For computational convenience, we define the integration measure of the scalar integrals in eq.~(\ref{eq:integralfamily}) as,
    \begin{eqnarray}
	\label{measure2}
	\widetilde{d^{d}k_{j}} \equiv 
    \frac{d^{d}k_{j}}{i \pi^{d/2}} \, \left(m_e^2\right)^{\epsilon}\frac{1}{\Gamma(1+\epsilon)} \ ,
\end{eqnarray}
such that the two tadpole MIs read,
\newcommand{\mymiii}{\bega \miii \ega}
\begin{equation}
\label{eq:tadpole_normalization}
\begin{split}
   \int
    \widetilde{d^{d}k_{1}}
    {1 \over D_1^2} 
     & = \bega \iloopmasterme{1.5} \ega 
= 
{ 1 \over \epsilon} \ , \qquad 
\int
    \widetilde{d^{d}k_{2}}
    {1 \over D_3^2} =
\bega \iloopmastermi{1.5} \ega 
= 
{ 1 \over \epsilon}\left(\frac{m_e^2}{m_i^2}\right)^\epsilon \ .
\end{split}
\end{equation}
The relation between the loop integral measure and the scalar integral measure, therefore is,
\begin{equation}\label{physicaltocanonicalmeasure}
\begin{split}
(\mu^2)^{\epsilon}
\frac{d^d k_j}{(2\pi)^{(d-2)}}
&=C_{\epsilon} \, \widetilde{d^{d}k_{j}},
\end{split}
\end{equation}
where $C_{\epsilon}$ is defined as
\begin{equation}
\label{eq:C_epsilon_def}
   C_{\epsilon}=\left(\frac{\mu^2}{m_e^2}\right)^{\epsilon}\left(\frac{i}{4}\,(4\pi)^{\epsilon}\,\Gamma(1+\epsilon)\right) \ ,
\end{equation}
with the limiting value
$
\lim_{\epsilon \to 0} C_{\epsilon} = i/4 \, .
$

\subsection{Master Integrals}

\begin{figure}[H]
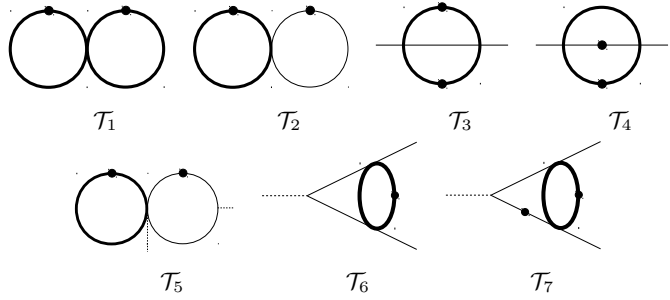

	\centering
	\captionsetup[subfigure]{labelformat=empty}
	\subfloat[\hspace{0.5cm}$\mathcal{T}_1$]{%
		\masteri{1.5}
	}
	\subfloat[\hspace{0.5cm}$\mathcal{T}_2$]{%
		\masterii{1.5}
	}
	\subfloat[\hspace{0.5cm}$\mathcal{T}_3$]{%
		\masteriii{1.5}
	}
	\subfloat[\hspace{0.5cm}$\mathcal{T}_4$]{%
		\masteriv{1.5}
	}\\
	\subfloat[\hspace{0.5cm}$\mathcal{T}_5$]{%
		\masterv{1.5}
	}
	\subfloat[\hspace{0.5cm}$\mathcal{T}_6$]{%
		\mastervi{1.5}
	}
	\subfloat[\hspace{0.5cm}$\mathcal{T}_7$]{%
		\mastervii{1.5}
	}
	\caption{Master integrals. Dots denote squared propagators.}
	\label{fig:tauMIs}
\end{figure}

Owing to the integration-by-parts relations \cite{Tkachov:1981wb,Chetyrkin:1981qh} and Laporta's algorithm \cite{Laporta:2000dsw}, 
all the integrals defined in eq.~(\ref{eq:integralfamily}) admit a decomposition in terms of $7$ master integrals, 
${\cal T}_i$ with $i=1,\ldots,7$, 
shown in Figure~\ref{fig:tauMIs}, obtained with the packages \textsc{LiteRed} \cite{Lee:2012cn,Lee_2014} and \textsc{Fire} \cite{Smirnov_2020} independently.

\subsubsection{System of Differential Equations}
The basis ${\mathcal{T}_i}$, up to $\epsilon$ rescalings, obeys a system of differential equations (sDEQ) whose matrix differential has a linear dependence on $\epsilon$. By using the method of the Magnus/Dyson exponential matrix~\cite{Argeri:2014qva}, this set is transformed into a canonical basis $\mathbf{I}$ with elements:
\begin{eqnarray}\label{eq:canonicalMIs}
\text{I}_1 &=& \epsilon^2 \mathcal{T}_1, \hspace{3cm} \text{I}_2= \epsilon^2 \mathcal{T}_2, \nonumber\\
\text{I}_3 & =& \epsilon ^2 m_e m_i \left(\mathcal{T}_3+2 \mathcal{T}_4\right), \qquad \text{I}_4 =  \epsilon ^2 m_e^2 \mathcal{T}_4, \nonumber \\
\text{I}_{5}&=& \epsilon ^2  \sqrt{-s}\sqrt{4 m_e^2-s} \mathcal{T}_5, \qquad \text{I}_6 = \epsilon ^3 \sqrt{-s}  \sqrt{4 m_e^2-s} \mathcal{T}_6, \nonumber \\
\text{I}_7 & = & -\frac{\epsilon^2}{2} \sqrt{-s}\sqrt{4 (m_e^2- m_i^2)-s} \left( 2(s-4 m^2_e) \mathcal{T}_7-2 \mathcal{T}_4-\mathcal{T}_3\right) \ ,
\end{eqnarray}
which obey a sDEQ having the following canonical structure~\cite{Henn:2013pwa}:
\begin{equation}
    d \mathbf{I}(\epsilon,x,y) = \epsilon \, d \mathbb{A}(x,y) \, \mathbf{I}(\epsilon,x,y), \qquad d = dx \, \frac{\partial }{\partial x} + dy \, \frac{\partial }{\partial y}.
\label{eq:canonical_DEQ}
\end{equation}
In this way, the dependence on the $\epsilon$ parameter is factorized, and the entries of the total differential matrix $d{\mathbb A}$ are rational in the variables $x$ and $y$.
The latter depend on the original variables $s, m^2_i$ and $m^2_e$, through the relations  
\begin{equation}
    - \frac{s}{m^2_e} = \frac{4 x^2}{1-x^2} \ , \qquad \frac{m_i}{m_e} = \frac{1-y^2}{1-2x y +y^2} \ ,
    \label{eq:change_vars_xy}
\end{equation}
with inverse
\begin{equation}
    x=\frac{\sqrt{-s}}{\sqrt{4 m^2_e -s}}\ , \qquad y= \frac{m_i \sqrt{-s} - m_e \sqrt{4(m^2_e-m^2_i)-s}}{(m_e+m_i) \sqrt{4 m^2_e-s}},
\end{equation}
whose expressions are obtained with the help of the package \textsc{RationalizeRoots} \cite{Besier_2020}. 
The special form of eq.~(\ref{eq:canonical_DEQ}) implies that the solution can be written as a Taylor series expansion in power of $\epsilon$, as 
\begin{equation}
    \mathbf{I}(\epsilon,x,y) = \sum_{j=0}^{\infty} \mathbf{I}^{(j)}(x,y) \, \epsilon^j\ ,
\end{equation}
with 
\begin{equation}
    \mathbf{I}^{(j)} = \sum_{i=0}^{j} \int_{\gamma} \underbrace{d \mathbb{A} \dots d \mathbb{A}}_{\text{$i$ times}}\,  \mathbf{I}^{(j-i)}(x_0,y_0)\ ,
\label{eq:solution_DEQ_implicit}
\end{equation}
where $\gamma$ is some regular path in the $(x,y)-$plane, and $\mathbf{I}^{(j-i)}(x_0,y_0)$ is a vector of boundary constants. In terms of the original variables, the boundary vector corresponds to the (canonical) MIs evaluated in the limit $s \to 0$ and $m_i=m_e$.\\
Explicitly, the matrix in eq.~(\ref{eq:canonical_DEQ}) is given by:
\begin{equation}
d \mathbb{A}(x,y)= \sum_{i=1}^{9} \mathbb{M}_i \, d \log \left( \eta_i(x,y) \right)
\label{eq:dA_exression}
\end{equation}
with 
\begin{equation}
\begin{split}
    \eta_1 & = y, \qquad \; \; \; \; \; \; \; \; \eta_2 = 1+x, \qquad  \; \; \; \; \; \; \eta_{3}=1+y,\\ 
     \eta_4 & = x-y, \qquad \; \;  \eta_5 =1-x, \qquad \; \; \; \; \; \; \eta_6 = 1-y,\\
   \eta_7 & = 1-xy, \qquad \eta_{8} = 1 - 2 x y + y^2, \; \; \eta_{9} = x - 2 y + x y^2 \ ,
\label{eq:alphabet}
\end{split}
\end{equation}
and where the coefficient matrices $\mathbb{M}_i$, shown in  Appendix~\ref{appendix:matrices_canonical}, have rational numbers as entries. \\
The solution to the differential equation is valid in the region
\begin{equation}\label{eq:solution_region}
    0<x<1 \quad \bigcap \quad 0<y < \frac{1 - \sqrt{1-x^2}}{x},
\end{equation}
where all the letters in eq.~(\ref{eq:alphabet}) are real and positive. In terms of the variables $s, m_i$ and $m_e$, eq.~(\ref{eq:solution_region}) corresponds to the (unphysical) region 
\begin{equation}
\label{eq:solution_region_smime}
 m_i>m_e \quad \bigcap \quad s< 4(m^2_e-m^2_i) \qquad  \left( m_e >0  \right).
\end{equation}
Given eqs.~(\ref{eq:dA_exression},\, \ref{eq:alphabet}), the solution of eq.~(\ref{eq:solution_DEQ_implicit}) can be written in terms of GPLs, defined recursively as
\begin{equation}
    G(a_1, \dots , a_n;z) = \int_{0}^{z} \frac{dt}{t-a_1} \, G(a_2, \dots, a_n;t), \qquad G(\underbrace{0, \dots, 0}_{n \text{ times}};z) = \frac{1}{n!} \, \log^n(z) \ .
\end{equation}

\subsubsection{Boundary Conditions}
The determination of the boundary constants proceeds by combining both quantitative and qualitative properties of the considered integrals:

\begin{itemize}
    \item the boundary values of $\text{I}_{1,2}$ are determined via direct integration, taking into account eq.~(\ref{eq:tadpole_normalization});
    \item the boundary constants of $\text{I}_{3,4}$ are determined by considering the subsystem formed by the first four MIs and using a different variable, $z=m_e/m_i$. The boundary constants for this subsystem are fixed at $z=0$. The boundary constants for $\text{I}_{3,4}$, expressed in the original variables $x$ and $y$, are determined by matching the general solution against the abovementioned subsystem, in the equal mass limit.
  \item the boundary values of $\text{I}_{5,6,7}$ are computed in the limit $s \to 0$ with $m_i=m_e$, where they are expected to vanish, due to the prefactors appearing in the definition of the canonical bases and the regularity of the MIs  $\mathcal{T}_{5,6,7}$.
\end{itemize}

As a useful consistency check, the boundary constants were also evaluated numerically with the package \textsc{AMFlow} \cite{Liu_2023} at the point $s=0,\,m_i=m_e$ and reconstructed using the \textsc{PSLQ} algorithm \cite{PSLQ}.

The boundary vector takes a simple form:
\begin{equation}
   \bold{I}(x_0,y_0)=\left(
       \begin{array}{c}
           1\\
           1\\
           \text{I}_3(x_0,y_0)\\
           \text{I}_4(x_0,y_0)\\
           0\\
           0\\
           0
       \end{array}
   \right)+\mathcal{O}(\epsilon^5),
\end{equation}
with
\begin{equation}
    \begin{aligned}
    \text{I}_3(x_0,y_0) &= -\frac{1}{4}\pi^2\epsilon^2+\frac{1}{4}\left(6 \pi ^2 \log (2)-21 \zeta (3)\right)\epsilon ^3\\
     &\quad+\frac{1}{120}\left(-4320 \text{Li}_4\left(\frac{1}{2}\right)+31 \pi ^4-180 \log ^4(2)-360 \pi ^2 \log ^2(2)\right)\epsilon ^4 \ , \\
    \text{I}_4(x_0,y_0) &= \frac{1}{3} \text{I}_3(x_0,y_0) \ .
    \end{aligned}
\end{equation}
We used \textsc{PolyLogTools}~\cite{Duhr_2019} for the algebraic manipulation of the GPLs, and \textsc{GiNaC}~\cite{Vollinga_2004sn} for their numerical evaluation.
The MIs were successfully compared against the numerical values provided by {\sc pySecDec}~\cite{Borowka_2018} and {\sc AMFlow}, as well as
against the set of MIs presented in \cite{Budassi:2021twh}.
 The analytic expression of the MIs, written in terms of GPLs up to weight $w=4$ are provided in the ancillary file {\tt <results.m>} accompanying this article, as well as the corresponding {\tt arXiv} version.
 \\

\section{Renormalization}\label{sec:renormalisation}

The diagrams depicted in Figure~\ref{fig:VkVertex} constitute a gauge invariant subset of vertex diagrams that depend on both $m_i$ and $m_e$, and thus the UV renormalization can be addressed independently of other contributions. The renormalization of other divergent graphs depending solely on a single mass scale $m_e$ has been performed separately in~\cite{Bonciani:2003ai}.
The renormalized vertex functions $\Gamma^{(k)\text{ren}}_{\mu}$ and $\Gamma^{(k)\text{ren}}$ are defined by the following combination of diagrams,
\begin{equation}\label{renormvertex}
\begin{aligned}
   \Gamma^{(k)\text{ren}}_{\mu}&=
    \vchb{\maintopologynolab{1.5}}
    +
    \vchb{\ilooptriangle{1.5}}
    +
    \vchb{\ziicountertermins{1.5}} \ , \\
    \Gamma^{(k)\text{ren}}&=
    \vchb{\maintopologynolabQCD{1.55}}
    +
    \vchb{\ilooptriangleQCD{1.65}}
    +
    \vchb{\ziicounterterminsQCD{1.65}} \ ,
\end{aligned}
\end{equation}
namely by adding two conuterterm diagrams to the unrenormalized vertices, and 
are computed in a two steps procedure.
The first type of counterterm diagrams represents the subtraction of the one-loop sub-divergence, achieved by renormalizing the vacuum polarisation insertion in the on-shell scheme, with
\begin{equation}
    \vchb{\ziiicountertermins{1.3}}=Z_3^{(1)}(-g_{\lambda\eta}\ell^2+\ell_\lambda \ell_\eta)\ ,
\end{equation}
where $\ell^{\mu}=k^{\mu}_1-p^{\mu}_1$ is the momentum flowing through the insertion. The one-loop renormalization constant $Z_3^{(1)}$ is implicitly defined by requiring 
\begin{equation}
        \lim_{\ell^2 \to 0}\left( \vchb{\iloopspinorsunrise{1}}+
        \vchb{\ziiicountertermins{1.2}}\right)=0 \ .
\end{equation}
The second type of counterterm diagrams, defined as,
\begin{equation}
    \vchb{\ziicountertermins{1.65}}=-i\left(Z_V^{(2)}g_{V}\gamma_\mu+Z_A^{(2)}g_{A}\gamma_{5}\gamma_{\mu}\right) \ ,
    \quad
    \vchb{\ziicounterterminsQCD{1.65}}=-i\left(Z_{S}^{(2)}g_S\mathbbm{1}+Z_{P}^{(2)}g_P\gamma_5\right) \ ,
\end{equation}
cancel the genuine two-loop residual divergences of the vertex.
The two-loop renormalization constants
with $Z^{(2)}_j,\ j\in \{V,A,S,P\}$, are implicitly defined by
\begin{equation}
\begin{aligned}
    \vchb{\ziicountertermins{1.65}}=&-\lim_{s\to0}\left[\vchb{\maintopologynolab{1.5}}
    +
    \vchb{\ilooptriangle{1.5}}\right] \ , \\
    \vchb{\ziicounterterminsQCD{1.65}}=&-\lim_{s\to0}\left[\vchb{\maintopologynolabQCD{1.55}}
    +
    \vchb{\ilooptriangleQCD{1.65}}\right] \ .
\end{aligned}
\end{equation}

Using IBPs, the renormalization constants admit the following expressions in terms of MIs,
\begin{equation}\label{eq:counterterms_expansions}
\begin{aligned}
    Z_3^{(1)}&=-\frac{4}{3}\left(\frac{\alpha}{\pi}\,C_{\epsilon}\right)\vchb{\iloopmastermi{1.2}}
     , \\
     Z_j^{(2)}& = 
       a_{j,1} \vchb{\masteri{1.2}}
     + a_{j,2} \vchb{\masterii{1.2}}
     + a_{j,3} \,\,\, \vchb{\masteriii{1.2}}
     + a_{j,4} \,\,\, \vchb{\masteriv{1.2}}  ,
\end{aligned}
\end{equation}
where the coefficients $a_{j,k}$ are not shown explicitly. 
After inserting the expression of the MIs, they can be expressed as Laurent series in $\epsilon$ as
\begin{equation}
    \begin{aligned}
        Z_3^{(1)}&= - \left( \frac{\alpha}{\pi}  C_{\epsilon} \right) \frac{4}{3 \epsilon} v^{-2  \epsilon}  \ , \\
        Z_V^{(2)}&=\left(\frac{\alpha}{\pi}\,C_{\epsilon}\right)^2\left(-\frac{1}{\epsilon}+\mathcal{O}\left(\epsilon^0\right)\right)\ , \\
        Z_A^{(2)}&=\left(\frac{\alpha}{\pi}\,C_{\epsilon}\right)^2\left(-\frac{1}{\epsilon}+\mathcal{O}\left(\epsilon^0\right)\right) \ , \\
        Z_S^{(2)}&=\left(\frac{\alpha}{\pi}\,C_{\epsilon}\right)^2\left(\frac{2}{\epsilon^2}-\frac{8}{3\epsilon}\left(1+3\log(v)\right)+\mathcal{O}\left(\epsilon^0\right)\right) \ , \\
        Z_P^{(2)}&=\left(\frac{\alpha}{\pi}\,C_{\epsilon}\right)^2\left(\frac{2}{\epsilon^2}-\frac{8}{3\epsilon}\left(1+3\log(v)\right)+\mathcal{O}\left(\epsilon^0\right)\right) \ ,
    \end{aligned}
\end{equation}
where $v=m_i/m_e$. 
We note that the difference between $Z_V$ and $Z_A$,
as well as between $Z_S$ and $Z_P$, begins from the finite term onwards.
More details on the evaluation of the renormalization constants can be found in Appendix~\ref{appendix:renorm}, where in particular the coefficients for $j=V$ are derived as an illustrative example.
\section{Results}\label{sec:results}

\subsection{Renormalized Form Factors}

For computational convenience, we introduce the rescaled form factors  $\mathcal{F}_{i}^{(k)\rm{ren}}$ and $\mathcal{G}_{i}^{(k)\rm{ren}}$, defined as
\begin{equation}
\label{eq:fk}
    F_{i}^{(k)\rm{ren}}  = C^2_{\epsilon} \,
   \mathcal{F}_{i}^{(k)\rm{ren}} \ , 
   \qquad
    G_{i}^{(k)\rm{ren}}  = C^2_{\epsilon} \,
    \mathcal{G}_{i}^{(k)\rm{ren}}  .
\end{equation}

The renormalized form factors $\mathcal{F}^{(k)\rm{ren}}_i $, 
with $i \in \{1,2,S,P\}$, and $\mathcal{G}^{(k)\rm{ren}}_i$, 
 with $i \in \{1,2\}$,
are expressed in terms of 74 
GPLs up to weight $w=3$, 
out of which 50 depending on $x$, with weights in the set 
\begin{equation}
    \left\{-1,1,y,\frac{1}{y},\frac{y^2+1}{2 y}\right\} \ ,
\end{equation}
and 24, on $y$, with weights in the set 
\begin{equation}
    \{-1,0,1,-i,i\} \ .
\end{equation}

Alternatively, we also provide their expression in terms of logarithms and classical polylogarithms, 
which turns out to be convenient for the numerical evaluation and for the series expansion of the form factors. 
The conversion of the GPLs into classical polylogarithms can be handled with the algorithm developed in \cite{Frellesvig:2016ske}~\footnote{ In particular, we observe that by exploiting the shuffle algebra of the GPLs and/or adding a small positive imaginary parameter $i \delta$ to suitable weights,
we were able to recast the results in terms of appropriate combinations of GPLs with complex weights, whose conversion to classical polylogarithms involves neither Heasviside $\theta$-function nor sgn-function in the region of interest.}. 
The renormalized form factors $\mathcal{F}^{(k)\mathrm{ren}}_{1,2}$ are in addition verified to be in numerical agreement with \cite{Fael:2018dmz}. Additionally, all the renormalized form factors are independently re-calculated using the variable choices and MIs presented in \cite{Budassi:2021twh}, and are found in complete agreement.

The expressions of the renormalized form factors $\mathcal{F}^{(k)\rm{ren}}_{1,2,S,P}$ and $\mathcal{G}^{(k)\rm{ren}}_{1,2}$ in terms of
GPLs, and in terms of classical polylogarithms constitute the main results of this communication.
Their expressions, too long to be shown here, as well as an implementation for their numerical evaluation, can be found in the ancillary file {\tt <results.m>}\,, and {\tt <evaluator.m>}, respectively accompanying this article, as well as the corresponding {\tt arXiv} version.

\subsection{Anomalous magnetic moment}
The limit $s \to 0$ of $F^{(k)\rm{ren}}_{2}$ corresponds to the two-loop, mass dependent contributions to the leptonic $g-2$. The limit can be considered at the diagrammatic level$-$following the discussion in appendix~\ref{appendix:renorm}$-$and the form factor can be expressed in terms of $\mathcal{T}_{1,2,3,4}$.

For generic $\epsilon$, its expression in terms of MIs reads
\begin{equation}
    F^{(k)\rm{ren}}_2(0,m_i,m_e)= a_1\,\,\vchb{\masteri{1.2}}+a_2\,\,\vchb{\masterii{1.2}}+ a_3\,\,\,\vchb{\masteriii{1.2}}+a_4 \,\,\, \vchb{\masteriv{1.2}},
\end{equation}
with
\begin{equation}
 a_i=\hat{a}_i C^{2}_{\epsilon},
\end{equation}
and
\begin{equation}
\begin{aligned}
    \hat{a}_1&=\frac{8 \epsilon  \left(\left(\epsilon  \left(\epsilon  \left(8 \epsilon ^3-4 \epsilon ^2+30 \epsilon +3\right)-41\right)+18\right) m_i^2-2 \left(\epsilon  \left(\epsilon  \left(8 \epsilon ^3-12 \epsilon -1\right)+4\right)+1\right) m_e^2\right)}{(\epsilon +1) (2 \epsilon -3) (2 \epsilon -1) (2 \epsilon +1) (3 \epsilon -2) (3 \epsilon -1) m_e^2} \ , \\
    \hat{a}_2&= \frac{8 \epsilon  \left((\epsilon +1) (\epsilon  (\epsilon  (12 \epsilon  (3 \epsilon -2)-55)+31)+6) m_e^2-3 (\epsilon  (\epsilon  (4 \epsilon  (5 \epsilon +6)-23)-29)+18) m_i^2\right)}{3 (\epsilon +1) (2 \epsilon -3) (2 \epsilon -1) (3 \epsilon -2) (3 \epsilon -1) m_e^2}\ , \\
    \hat{a}_3&=\frac{8 m_i^2 \left((\epsilon  (\epsilon  (4 (\epsilon -1) \epsilon  (4 \epsilon +3)-13)+5)+6) m_e^2+\left(\epsilon  \left(\epsilon  \left(3-2 \epsilon  \left(4 \epsilon ^2-6 \epsilon +27\right)\right)+67\right)-30\right) m_i^2\right)}{(\epsilon +1) (2 \epsilon -3) (2 \epsilon -1) (3 \epsilon -2) (3 \epsilon -1) m_e^2} \ , \\
    \hat{a}_4&=\frac{16 \left((\epsilon +1)^2 (\epsilon  (4 \epsilon  (5 \epsilon -6)-1)+6) m_e^2 m_i^2-4 (\epsilon -1) \epsilon  (\epsilon +1) \left(4 \epsilon ^2-2 \epsilon -1\right) m_e^4\right)}{(\epsilon +1) (2 \epsilon -3) (2 \epsilon -1) (3 \epsilon -2) (3 \epsilon -1) m_e^2}\\
    &\quad\,\,+\frac{(\epsilon  (\epsilon  (23-4 \epsilon  (5 \epsilon +6))+29)-18) m_i^4}{(\epsilon +1) (2 \epsilon -3) (2 \epsilon -1) (3 \epsilon -2) (3 \epsilon -1) m_e^2}\ . \\
\end{aligned}
\label{eq:gm2_fulleps}
\end{equation}
Inserting the explicit expressions for the MIs, and considering 
just the finite term in the $\epsilon \to 0$ limit, eq.~(\ref{eq:gm2_fulleps}) yields,
\begin{equation}
\begin{split}
    F_2^{(k)\mathrm{ren}}(0,m_i,m_e)= &  \frac{4}{z^2}-\frac{25}{36} +\left(\frac{1}{3}-\frac{3}{z^2}\right)
   G(0;z)+\frac{\left(4+5 z -z^3\right)
   }{2 z^4} G(-1,0;z)\\
   & +\frac{\left(4 -5 z + z^3\right)
   }{2 z^4} G(1,0;z),
    \label{eq:gm2}
\end{split}
\end{equation}
where $z=m_e/m_i$.\\
Eq.~(\ref{eq:gm2}) is an alternative, yet equivalent expression to the one presented in~\cite{Passera:2004bj}\footnote{The variable $x$ in eq.~(5) in ref.~\cite{Passera:2004bj} corresponds to $z^{-1}$.} and revisited in \cite{Henn_2021}.
It can be written in terms of logarithms and dilogarithms, upon the substitutions
\begin{align}
    & G(0;z)=\log(z), \nonumber \\
    & G(1,0;z)=\log(z)\,\log(1-z)+\mathrm{Li}_2(z), \nonumber \\
    & G(-1,0;z)=\log(z)\,\log(1+z)+\mathrm{Li}_2(-z). 
\end{align}

\section{Conclusions}\label{sec:conclusion}

We presented the analytic evaluation of the second-order corrections to the massive form factors, coming from two-loop vertex diagrams with a vacuum polarization insertion, with exact dependence on the 
external and internal fermion masses, and on the squared momentum transfer.
We considered vector, axial-vector, scalar, and pseudoscalar interactions in the coupling between the external fermion and the external field.
The calculation was performed within the dimensional regularization scheme. Using integration-by-parts identities, the form factors were decomposed in terms of a basis of seven master integrals. The latter were evaluated by means of the differential equation method, making use of  Magnus exponential matrix. 
The renormalized form factors were expressed in terms of generalised polylogarithms up to weight three, and in addition converted to classical polylogarithms. 

The presented results can be considered as the last, missing contributions to the problem of the analytic evaluation of the second-order corrections to the massive form factors in QED and QCD, within the dimensional regularisation scheme, a problem which began to be addressed about two decades ago.
The expressions of the form factors evaluated in this work can be straightforwardly applied in the context of the evaluation of the next-to-next-to-leading order virtual QED and QCD corrections 
to the decay of a massive neutral boson into heavy particles, or to the 
four (massive) fermion scattering amplitudes. \\

\subsection*{Acknowledgements}
We thank Carlo Carloni Calame and Manoj Mandal for interesting discussions and for fostering our collaboration, as well as  
Yannick Ulrich, Tim Engel and Adrian Signer, for the numerical comparisons of the renormalised vector form factors. 
It is a pleasure to acknowledge the colleagues of the theory initiative of the MUonE collaboration for stimulating discussions at various stages.
The work of TA was supported by Deutsche Forschungsgemeinschaft
(DFG) through the Research Unit FOR 2926, \textit{Next Generation perturbative QCD for Hadron Structure:
Preparing for the Electron-ion collider}, project number 409651613. FG has been supported by the Cluster of Excellence \textit{Precision Physics, Fundamental Interactions, and Structure of Matter} (PRISMA EXC 2118/1) funded by the German Research Foundation (DFG) within the German Excellence Strategy (Project ID 390831469).\\

\newpage
\appendix
\section{Matrices for the Canonical Differential Equation}
\label{appendix:matrices_canonical}
In this appendix we list the matrices $\{\mathbb{M}_i\}_{i=1}^9$ appearing in eq.~(\ref{eq:dA_exression})
\begin{equation}
\begin{split}
& \mathbb{M}_1=  \left(
\begin{array}{ccccccc}
 0 & 0 & 0 & 0 & 0 & 0 & 0 \\
 0 & 0 & 0 & 0 & 0 & 0 & 0 \\
 -\frac{1}{2} & \frac{1}{2} & 1 &
   -3 & 0 & 0 & 0 \\
 -\frac{1}{2} & \frac{1}{2} & 1 &
   -3 & 0 & 0 & 0 \\
 0 & 0 & 0 & 0 & 0 & 0 & 0 \\
 0 & 0 & 0 & 0 & 0 & 0 & 0 \\
 1 & -1 & -2 & 6 & 0 & 0 & 0 \\
\end{array}
\right),
\quad
\mathbb{M}_2=\left(
\begin{array}{ccccccc}
 0 & 0 & 0 & 0 & 0 & 0 & 0 \\
 0 & 0 & 0 & 0 & 0 & 0 & 0 \\
 0 & 0 & 0 & 0 & 0 & 0 & 0 \\
 0 & 0 & 0 & 0 & 0 & 0 & 0 \\
 0 & -1 & 0 & 0 & 1 & 0 & 0 \\
 0 & 0 & 0 & 2 & 0 & -1 & 1 \\
 1 & -1 & 0 & 6 & 1 & -3 & 3 \\
\end{array}
\right),\quad
\mathbb{M}_3=\left(
\begin{array}{ccccccc}
 -4 & 0 & 0 & 0 & 0 & 0 & 0 \\
 0 & -2 & 0 & 0 & 0 & 0 & 0 \\
 0 & 0 & -4 & 0 & 0 & 0 & 0 \\
 1 & -1 & 0 & 0 & 0 & 0 & 0 \\
 0 & 0 & 0 & 0 & -2 & 0 & 0 \\
 0 & 0 & 0 & 0 & 0 & -6 & -2 \\
 0 & 0 & 0 & 0 & -2 & 6 & 2 \\
\end{array}
\right),\\
& \mathbb{M}_4 = \left(
\begin{array}{ccccccc}
 0 & 0 & 0 & 0 & 0 & 0 & 0 \\
 0 & 0 & 0 & 0 & 0 & 0 & 0 \\
 -\frac{1}{2} & \frac{1}{2} & 1 &
   -3 & 0 & 0 & 0 \\
 -\frac{1}{2} & \frac{1}{2} & 1 &
   -3 & 0 & 0 & 0 \\
 0 & 0 & 0 & 0 & 0 & 0 & 0 \\
 0 & 0 & 0 & 0 & 0 & 0 & 0 \\
 -1 & 1 & 2 & -6 & 0 & 0 & 0 \\
\end{array}
\right),\quad
\mathbb{M}_5=\left(
\begin{array}{ccccccc}
 0 & 0 & 0 & 0 & 0 & 0 & 0 \\
 0 & 0 & 0 & 0 & 0 & 0 & 0 \\
 0 & 0 & 0 & 0 & 0 & 0 & 0 \\
 0 & 0 & 0 & 0 & 0 & 0 & 0 \\
 0 & 1 & 0 & 0 & 1 & 0 & 0 \\
 0 & 0 & 0 & -2 & 0 & -1 & -1 \\
 1 & -1 & 0 & 6 & -1 & 3 & 3 \\
\end{array}
\right),\quad
\mathbb{M}_6 = \left(
\begin{array}{ccccccc}
 -4 & 0 & 0 & 0 & 0 & 0 & 0 \\
 0 & -2 & 0 & 0 & 0 & 0 & 0 \\
 0 & 0 & -4 & 0 & 0 & 0 & 0 \\
 1 & -1 & 0 & 0 & 0 & 0 & 0 \\
 0 & 0 & 0 & 0 & -2 & 0 & 0 \\
 0 & 0 & 0 & 0 & 0 & -6 & 2 \\
 0 & 0 & 0 & 0 & 2 & -6 & 2 \\
\end{array}
\right),\nonumber\\
& \mathbb{M}_7 =\left(
\begin{array}{ccccccc}
 0 & 0 & 0 & 0 & 0 & 0 & 0 \\
 0 & 0 & 0 & 0 & 0 & 0 & 0 \\
 \frac{1}{2} & -\frac{1}{2} & 1 & 3
   & 0 & 0 & 0 \\
 -\frac{1}{2} & \frac{1}{2} & -1 &
   -3 & 0 & 0 & 0 \\
 0 & 0 & 0 & 0 & 0 & 0 & 0 \\
 0 & 0 & 0 & 0 & 0 & 0 & 0 \\
 -1 & 1 & -2 & -6 & 0 & 0 & 0 \\
\end{array}
\right),\quad
\mathbb{M}_8=\left(
\begin{array}{ccccccc}
 4 & 0 & 0 & 0 & 0 & 0 & 0 \\
 0 & 2 & 0 & 0 & 0 & 0 & 0 \\
 0 & 0 & 2 & 0 & 0 & 0 & 0 \\
 0 & 0 & 0 & 6 & 0 & 0 & 0 \\
 0 & 0 & 0 & 0 & 2 & 0 & 0 \\
 0 & 0 & 0 & 0 & 0 & 6 & 0 \\
 0 & 0 & 0 & 0 & 0 & 0 & 2 \\
\end{array}
\right), \quad
\mathbb{M}_9=\left(
\begin{array}{ccccccc}
 0 & 0 & 0 & 0 & 0 & 0 & 0 \\
 0 & 0 & 0 & 0 & 0 & 0 & 0 \\
 0 & 0 & 0 & 0 & 0 & 0 & 0 \\
 0 & 0 & 0 & 0 & 0 & 0 & 0 \\
 0 & 0 & 0 & 0 & 0 & 0 & 0 \\
 0 & 0 & 0 & 0 & 0 & 0 & 0 \\
 0 & 0 & 0 & 0 & 0 & 0 & -4 \\
\end{array}
\right).
\end{split}
\end{equation}

\section{Evaluating the Renormalization Constants}
\label{appendix:renorm}

In this appendix, we describe in detail the evaluation of the renormalization constants for the vector form factors.
A similar procedure has been followed in the other cases.

\subsection{ $Z_3^{(1)}$ Renormalization Constant}\label{subsec:z31 renorm constant}
The renormalization constant $Z_3^{(1)}$ is defined implicitly by the following equation
\begin{equation}
        \lim_{\ell^2 \to 0}\left( \vchb{\iloopspinorsunrise{1}}+
        \vchb{\ziiicountertermins{1.2}}\right)=0.
\end{equation}
In order to derive its explicit expression, we expand the one-loop two point function in terms of scalar master integrals
\begin{equation}
\label{1loopformfactor}
\begin{split}
\vchb{\iloopspinorsunrise{1}}
        &=
        \left(\frac{4  \left(-2 m_i^2+\ell^2( \epsilon -1)\right)}{  \ell^2 (2 \epsilon -3)} \, \, 
        \vchb{\iloopscalarsunrise{1.2}}
        +\frac{8  m_i^2}{\ell^2 (2 \epsilon -3)}  \, \, 
        \vchb{\iloopmastermi{1.2}}
     \right)\\
&\,\times \left(\frac{\alpha}{\pi}\,C_{\epsilon}\right)(- g_{\mu \nu}\ell^2+\ell_\mu \ell_\nu ) \ .
\end{split}
\end{equation}
To evaluate the limit $\ell^2\to0$ it is necessary to expand the two point function, as a power series in $\ell^2$, up to the first order as:
\begin{equation}\label{series expansion}
    \vchb{\iloopscalarsunrise{1.2}}=\vchb{\iloopmastermi{1.2}}\left(j_0+j_1 \ell^2+\mathcal{O}(\ell^4)\right) \ ,
\end{equation}
so that, by using Taylor series expansion, we can identify  
\begin{equation}
\begin{aligned}
j_0\vchb{\iloopmastermi{1.2}}=\left.\vchb{\iloopscalarsunrise{1.2}}\right\rvert_{\ell^2=0}\ ,\qquad
j_1\vchb{\iloopmastermi{1.2}}=\frac{d}{d\ell^2}\left.\left(\vchb{\iloopscalarsunrise{1.2}}\right)\right\rvert_{\ell^2=0}\ .
\end{aligned}
\end{equation}

By direct inspection, the $\ell^2 \to 0$ limit can be taken diagrammatically,
(as in this example the integral is finite),
\begin{equation}
    \left.\vchb{\iloopscalarsunrise{1.2}}\right\rvert_{\ell^2=0}=\vchb{\iloopmastermi{1.2}} \ ,
\end{equation}
which implies the value $j_0=1$. \\ 
Using IBPs, we can also evaluate the first derivative of the 2-point integral,
\begin{equation}\label{derivative}
\frac{d}{d\ell^2}\left(\vchb{\iloopscalarsunrise{1.2}}\right)=\frac{2m_i^2}{\ell^2(4m_i^2-\ell^2)}\vchb{\iloopscalarsunrise{1.2}}-\frac{\ell^2 \epsilon -2 m_i^2}{\ell^2(\ell^2-4 m_i^2)}\vchb{\iloopmastermi{1.2}} \ ,
\end{equation}
(which corresponds to the differential equation), 
and take the $\ell^2 \to 0$ limit as follows,
\begin{equation}
\begin{split}
    j_1 \ \vchb{\iloopmastermi{1.2}} 
    &=\left.\left(\frac{2m_i^2}{\ell^2(4m_i^2-\ell^2)}\vchb{\iloopscalarsunrise{1.2}}-\frac{\ell^2 \epsilon -2 m_i^2}{\ell^2(\ell^2-4 m_i^2)}\vchb{\iloopmastermi{1.2}}\right)\right\rvert_{\ell^2=0}\ ,\\
    &=\left.\left(\frac{2m_i^2}{\ell^2(4m_i^2-\ell^2)}\vchb{\iloopmastermi{1.2}}\left(1+j_1 \ell^2+\mathcal{O}(\ell^4)\right)-\frac{\ell^2 \epsilon -2 m_i^2}{\ell^2(\ell^2-4 m_i^2)}\vchb{\iloopmastermi{1.2}}\right)\right\rvert_{\ell^2=0}\ ,
\end{split}
\end{equation}
which simplifies to
\begin{equation}
    j_1=\left.\frac{1}{4} \left(\frac{\epsilon }{m_i^2}-2 j_1+\mathcal{O}(\ell^2)\right)\right\rvert_{\ell^2=0}=\frac{1}{4} \left(\frac{\epsilon }{m_i^2}-2 j_1\right) \ .
\end{equation}
The latter can be read as an equation in $j_1$,
whose solution gives $j_1=\epsilon/6m_i^2$, hence fixing our Taylor series to be
\begin{equation}
   \label{powseriesfixed}
    \vchb{\iloopscalarsunrise{1.2}}=\vchb{\iloopmastermi{1.2}}\left(1+\frac{\epsilon}{6m_i^2} \ell^2
    \right) +\mathcal{O}(\ell^4) \ .
\end{equation}
Finally, this result can be inserted into eq.~(\ref{1loopformfactor}), to obtain
\begin{equation}
    \vchb{\iloopspinorsunrise{1}}=\left(\frac{4}{3}\vchb{\iloopmastermi{1.2}}+\mathcal{O}(\ell^2)\right)\left(\frac{\alpha}{\pi}\,C_{\epsilon}\right)(- g_{\mu \nu}\ell^2+\ell_\mu \ell_\nu ) \ ,
\end{equation}
and thus
\begin{equation}
    Z_3^{(1)}=-\frac{4}{3}\left(\frac{\alpha}{\pi}\,C_{\epsilon}\right)\vchb{\iloopmastermi{1.2}}=-\frac{4}{3\epsilon}\left(\frac{m_e^2}{m_i^2}\right)^{\epsilon}\left(\frac{\alpha}{\pi}\,C_{\epsilon}\right) \ .
\end{equation}
where in the last equality we have used eq.~(\ref{eq:tadpole_normalization}).
\subsection{Diagram for Subdivergence Renormalization}
\label{sec:1loopdiagwithcountertermQED}
We consider the decomposition in terms of MIs of
\begin{eqnarray}
\label{ctdiagexpandedscalar}
    \text{Tr}\left[ P_{1}^\mu
    \left(\vchb{\ilooptriangle{1.2}}\right)
    \right] &=& 
    \left(\frac{\alpha}{\pi}\,C_{\epsilon}\right)
    \left(
    \frac{2  \left(2 \left(\epsilon -1\right) \left(2
   \epsilon +1\right) m_e^2+s\right)}{  \epsilon  (2 \epsilon -1) \left(4 m_e^2-s\right)}\vchb{\iloopmasterme{1.2}}\right. \nonumber \\
    && \qquad \qquad 
    \left.- 
    \frac{  \left(4 m_e^2-s \left(2 \epsilon ^2-\epsilon
   +2\right)\right)}{  \epsilon  \left(4 m_e^2-s\right)}\vchb{\iloopscalarsunrise{1.2}}\right)Z_3^{(1)} \ ,
\end{eqnarray}
and

\begin{equation}
\begin{aligned}
\label{ctdiagexpandedscalar2}
    \text{Tr}\left[ P^{\mu}_{2}
    \left(\vchb{\ilooptriangle{1.2}}\right)
    \right] &= 
    \left(\frac{\alpha}{\pi}\,C_{\epsilon}\right)
    \left(
    \frac{4 (2 \epsilon +1) m_e^2}{(2 \epsilon -1) \left(s-4 m_e^2\right)}
    \vchb{\iloopmasterme{1.2}}\right. \nonumber \\
    & \qquad \qquad \quad \,\,\,\,
    \left.- 
    \frac{4 (2 \epsilon +1) m_e^2}{s-4 m_e^2}
    \vchb{\iloopscalarsunrise{1.2}}\right)Z_3^{(1)} \ .
\end{aligned}
\end{equation}
The one-loop MIs, although simple, are computed with the method of differential equations, using the change of variables as in eq.~(\ref{eq:change_vars_xy}) to ensure compatibility with the unrenormalized form factors.

\subsubsection{ $Z_V^{(2)}$ Renormalization Constant}
\label{z31eval}
We define $Z_V^{(2)}$ through the requirement that
\begin{equation}\label{z12 def}
\begin{gathered}
    \lim_{s\to0}\text{Tr}\left( P_{F1}^\mu
    \,\Gamma^{(k)\text{ren}}_{\mu}
    \right)=0
\end{gathered} \ ,
\end{equation}
which, by employing eq.~(\ref{renormvertex}), implies 
\begin{equation}\label{z12 def simp}
\begin{aligned}
    Z_V^{(2)}&=
    -\lim_{s\to0}\text{Tr}\left[ P_{F1}^\mu \,
    \left(\vchb{\ilooptriangle{1.2}}\right)
    \right]-\lim_{s\to0}\text{Tr}\left[ P_{F1}^\mu
    \left(\vchb{\maintopologynolab{1.2}}\right)
    \right] \\
    &=-\lim_{s\to0}\text{Tr}\left[ P_{F1}^\mu \,
    \left(\vchb{\ilooptriangle{1.2}}\right)
    \right]-\lim_{s\to0}\left[ F_1^{(k)} \right] \ .
\end{aligned}
\end{equation}
The first contribution on the r.h.s. can be evaluated 
by using eq.~(\ref{ctdiagexpandedscalar}) and eq.~(\ref{powseriesfixed}), giving 
\begin{eqnarray}
      \lim_{s\to0}\text{Tr}\left[ P_{F1}^\mu
    \left(\vchb{\ilooptriangle{1.2}}\right)
    \right] &=& 
    \left(\frac{\alpha}{\pi}\,C_{\epsilon}\right)\frac{  (2 \epsilon -3)}{  (2 \epsilon -1)}\,Z_3^{(1)}\vchb{\iloopmasterme{1.2}} 
    \nonumber \\ 
    &=&\left(\frac{\alpha}{\pi}\,C_{\epsilon}\right)^2\frac{4 (3-2 \epsilon )}{3 (2 \epsilon -1)}\vchb{\masterii{1}}.  
\end{eqnarray}
To evaluate $F_1^{(k)}$ in the limit $s\to0$ we proceed in a similar manner to in that in Section \ref{subsec:z31 renorm constant}. 
By considering the leading term of the master integrals $\mathcal{T}_{5,6,7}$ with respect to $s$ we obtain
\begin{equation}
\begin{aligned}
    \vchb{\masterv{1.2}}&=\vchb{\masterdotsii{1.1}}+\mathcal{O}(s)=-\frac{2\epsilon}{4m_e^2}\vchb{\masterii{1.1}}+\mathcal{O}(s) \ , \\
    \vchb{\mastervi{1.2}}&=\vchb{\masteriv{1.1}}+\mathcal{O}(s) \ , \\
    \vchb{\mastervii{1.2}}&=\vchb{\masterdotsiv{1.1}}+\mathcal{O}(s)=\frac{\epsilon }{8 m_e^2(m_e^2-  m_i^2)}\left(\vchb{\masteri{1.1}}+\vchb{\masterii{1.1}}\right)\\
    &\qquad\qquad+\frac{m_i^2(2 \epsilon +1)-m_e^2}{8 m_e^2 \left(m_e^2-m_i^2\right)}\vchb{\masteriii{1.1}}+\frac{(2 \epsilon +1) m_i^2-(3 \epsilon +1) m_e^2}{4 m_e^2 \left(m_e^2-m_i^2\right)}\vchb{\masteriv{1.1}}+\mathcal{O}(s) \ . \\
\end{aligned}
\end{equation}
The integrals $\mathcal{T}_{1,2,3,4}$ do not depend on $s$ and thus do not need to be expanded. Using these identities we can evaluate the limit as
\begin{equation}
\begin{gathered}
\lim_{s\to0}\left[F_1^{(k)}\right]=a_1 \vchb{\masteri{1.2}}+ a_2 \vchb{\masterii{1.2}}+a_3 \,\,\, \vchb{\masteriii{1.2}}+a_4 \,\,\, \vchb{\masteriv{1.2}} \ ,
\end{gathered}
\end{equation}
with
\begin{equation}
    a_i=\hat{a}_i\,C_{\epsilon}^2 \ , \nonumber
\end{equation}
and
\begin{equation}
\begin{aligned}
    \hat{a}_1 & =\left(\frac{2 \epsilon  (4 \epsilon -5) (\epsilon  (2 \epsilon +7)-7) (\epsilon +1)^2}{(\epsilon -1) (2 \epsilon +1) (2 \epsilon +3) \left(9 \epsilon ^3-7 \epsilon +2\right)}+\frac{4 m_i^2 \epsilon  (\epsilon  ((5-4 \epsilon ) \epsilon -7)+4)}{m_e^2 (2 \epsilon -1) \left(9 \epsilon ^3-7 \epsilon +2\right)}\right) \ ,\\
    \hat{a}_2& =\left(\frac{4 m_i^2 \epsilon  (\epsilon  (\epsilon  (2 \epsilon +19)-7)-4)}{m_e^2 (\epsilon +1) (2 \epsilon -1) (3 \epsilon -2) (3 \epsilon -1)}-\frac{8 (\epsilon -1) (\epsilon +1)}{(3 \epsilon -2) (3 \epsilon -1)}\right) \ ,\\
    \hat{a}_3&=\left(\frac{4 m_i^4 (\epsilon  (\epsilon  (2 \epsilon  (4 \epsilon -7)+17)+9)-10)}{ m_e^2 (\epsilon +1) (2 \epsilon -1) (3 \epsilon -2) (3 \epsilon -1)}+\frac{4 m_i^2 (\epsilon  (\epsilon  (21-4 \epsilon  (\epsilon +3))+3)-6)}{(\epsilon +1) (2 \epsilon -1) (3 \epsilon -2) (3 \epsilon -1)}\right) \ ,\\
    \hat{a}_4&=\left(\frac{8 m_i^4 (\epsilon  (\epsilon  (2 \epsilon +19)-7)-4)}{ m_e^2 (\epsilon +1) (2 \epsilon -1) (3 \epsilon -2) (3 \epsilon -1)}+\frac{8 m_i^2 (-\epsilon -1) (\epsilon  (\epsilon  (2 \epsilon +15)-21)+6)}{(\epsilon +1) (2 \epsilon -1) (3 \epsilon -2) (3 \epsilon -1)}\right.\\ 
    &\qquad\left.+\frac{16 m_e^2 (\epsilon -1) (\epsilon +1)}{(3 \epsilon -2) (3 \epsilon -1)}\right) \ . \\
\end{aligned}
\label{eq:c_hat_2L_trace}
\end{equation}

Finally, by summing the two relevant contributions, the expression of $Z_V^{(2)}$ reads
\begin{equation}
\begin{gathered}
    Z_V^{(2)}=-\left(\frac{\alpha}{\pi}\right)^2\left(a_1 \vchb{\masteri{1.2}}+ \left[a_2+\frac{4(3-2\epsilon)}{3(2\epsilon-1)}C_{\epsilon}^2\right] \vchb{\masterii{1.2}}+a_3 \, \vchb{\masteriii{1.2}}+a_4 \, \vchb{\masteriv{1.2}}\right) \ .
\end{gathered}
\end{equation}

By substituting in the relevant expansions for the master integrals, at leading order, the expression for  $Z_V^{(2)}$ takes the simple form
\begin{equation}
    Z_V^{(2)}=\left(\frac{\alpha}{\pi}\,C_{\epsilon}\right)^2\left(-\frac{1}{\epsilon}+\mathcal{O}(\epsilon^0)\right).
\end{equation}

\bibliographystyle{JHEP}
\bibliography{biblio}
\end{document}